\documentclass[times,twocolumn,final,authoryear]{elsarticle}

\usepackage{jasr}
\usepackage{framed,multirow}

\usepackage{amssymb}
\usepackage{latexsym}
\usepackage{amssymb}

\usepackage[switch]{lineno}

\usepackage{url}
\usepackage{xcolor}
\definecolor{newcolor}{rgb}{.8,.349,.1}

\usepackage[citebordercolor=white]{hyperref}
\hypersetup{
    colorlinks=true,
    linkcolor=blue
    }

\journal{Advances in Space Research}

\begin{document}

\verso{Given-name Surname \textit{etal}}

\begin{frontmatter}

\title{Sun CubE OnE: A Multi-wavelength Synoptic Solar Micro Satellite
}

\author[1]{L. Giovannelli}
\author[1]{F. Berrilli \corref{cor1}}
\cortext[cor1]{Corresponding author:}
\ead{francesco.berrilli@roma2.infn.it}
\author[2]{M. Casolino}
\author[3]{F. Curti}
\author[1]{D. Del Moro}
\author[1]{D. Calchetti}
\author[1]{M. Cantoresi}
\author[3]{A. D'Ambrosio}
\author[1]{G. Francisco}
\author[1]{P. Giobbi}
\author[2]{L. Marcelli}
\author[1]{P. Mazzotta}
\author[1]{R. Mugatwala}
\author[1]{G. Pucacco}
\author[1]{R. Reda}
\author[1]{S. K Dhara}
\author[1]{F. Tombesi}
\author[6]{D. Blandino}
\author[4]{N. Benigno}
\author[6]{M. Cilia}
\author[5]{A. Di Salvo}
\author[4]{V. Di Tana}
\author[4]{F. Ingiosi}
\author[5]{S. Loddo}
\author[6]{M. Marmonti}
\author[6]{M. Musazzi}
\author[4]{S. Simonetti}
\author[4]{G. Truscelli}


\address[1]{Department of Physics, University of Rome Tor Vergata, Via della Ricerca Scientifica 1, Rome, 00133}
\address[2]{INFN, Structure of Rome Tor Vergata, Via della Ricerca Scientifica 1, Rome, 00133, Italy}
\address[3]{Sapienza University of Rome, School of Aerospace Engineering - ARCAlab, Via Salaria 851, 00138 Rome, Italy}
\address[4]{ARGOTEC S.r.l., Via Cervino 52, 10155 Turin, Italy}
\address[5]{NEXT Ingegneria dei Sistemi S.p.A., Via Giacomo Peroni 452, 00131 Roma, Italy}
\address[6]{Optec S.p.A., Via Mantegna 34, 20015 Parabiago (MI) - Italy}

\received{}
\finalform{}
\accepted{}
\availableonline{}
\communicated{}

\begin{abstract}
The Sun cubE onE (SEE) is a 12U CubeSat mission proposed for a phase A/B study to the Italian Space Agency that will investigate Gamma and X-ray fluxes and ultraviolet (UV) solar emission to support studies in Sun-Earth interaction and Space Weather from LEO. More in detail, SEE's primary goals are to measure the flares emission from soft-X to Gamma ray energy range and to monitor the solar activity in the Fraunhofer Mg II doublet at 280 nm, taking advantage of a full disk imager payload. 
The Gamma and X-ray fluxes will be studied with unprecedented temporal resolution and with a multi-wavelength approach thanks to the combined use of silicon photodiode and silicon photomultiplier (SiPM) -based detectors. The flare spectrum will be explored from the keV to the MeV range of energies by the same payload, and with a cadence up to 10 kHz and with single-photon detection capabilities to unveil the sources of the solar flares. The energy range covers the same bands used by GOES satellites, which are the standard bands for flare magnitude definition. At the same time SiPM detectors combined with scintillators allow to cover the non-thermal bremsstrahlung emission in the gamma energy range.
Given its UV imaging capabilities, SEE will be a key space asset to support detailed studies on solar activity, especially in relation to ultraviolet radiation which strongly interacts with the upper layers of the Earth's atmosphere, and in relation to space safety, included in the field of human space exploration. The main goal for the UV payload is to study the evolution of the solar UV emission in the Mg II band at two different time scales: yearly variations along the solar cycle and transient variations during flare events. The Mg II index is commonly used as a proxy of the solar activity in the Sun-as-a-star paradigm, in which solar irradiance variations in the UV correlate with the variations in stratospheric ozone concentrations and other physical parameters of the Earth high atmosphere.
SEE data will be used together with space and ground-based observatories that provide Solar data  (e.g. Solar Orbiter, IRIS, GONG, TSST), high energy particle fluxes (e.g. GOES, MAXI, CSES) and geomagnetic data in a multi-instrument/multi-wavelength/multi-messenger approach.

\end{abstract}


\end{frontmatter}


\section{Introduction}
\label{sec1}

As we are entering an era of space exploration dominated by smaller satellites and vast constellation, science missions are moving more and more towards the exploitation of CubeSat standards \citep[see e.g.][]{cubesat}.
Furthermore, the revamp of deep space human exploration on the Moon and lunar orbit and future plans for Mars exploration places solar and space weather missions in the next decade at high priority.
The Sun cubE onE (SEE) is a mission planned in low Earth orbit (LEO) to study Sun-Earth interaction exploiting a 12U CubeSat platform and Commercial-off-the-shelf (COTS) approach for fast development and relative lower budget compared to even small-class missions.
SEE primary goal is to study the high energy solar irradiance variations, from UV to gamma rays, exploring time cadences that vary from the unexplored sub-seconds in the solar flares to the yearly time scales. Furthermore, SEE mission wants to provide a space weather platform for the space safety stakeholders.
To fulfil those goals SEE mission will include two primary scientific payloads: an UV imager in the 280 nm band and a gamma and X-ray instrument for high cadence spectroscopy during flare emission. Furthermore, once SEE capabilities will be proved in-orbit, it will pave the way for a SEE-based series of low-cost cubesats, to provide longer time coverage over a full-solar cycle. A launch every 1.5 years would provide redundancy, data cross-calibration complete time coverage over the years and would transform SEE from a proof-of-concept cubesat mission to a space weather asset based on a series of cubesats.

The solar spectral irradiance (SSI) in the UV/X-ray/gamma bands can only be measured by instruments aboard space missions. Solar variability in UV and high energy regions of the electromagnetic spectrum is far more important than what happens in integrated power at 1 A.U. (the total solar irradiance or TSI). In fact, the typical variations of the TSI during periods of higher solar activity, characterized by peaks in the number of facular regions and sunspots in the solar photosphere and by a higher frequency of flares and coronal mass ejections (CME), increases by only about one tenth of a percent (or about 2 Wm$^{-2}$ between the maximum and minimum of an 11-year solar cycle). But, relative changes of TSI are not evenly distributed across the solar spectrum: they are concentrated in the UV, so that wavelengths less than 400 nm contribute about 9\% to the change, and about 32\% of the radiation variation over a solar cycle occurs below 250 nm \citep{lean1989,snow2010,ermolli2013}. 

In particular, we propose to study the variability of the solar emission in the Mg II doublet at 280 nm with a UV imager payload, which would be able to explore the chromospheric structures producing such emissions.
This doublet is particularly relevant because the Mg II index (core-to-wing ratio) is based on the emission profile of Mg II doublet and it is usually used as a proxy for the spectral solar irradiance variability from the UV to EUV associated with the periods of 11-year (solar cycle) and about 27 days (solar rotation). The index is computed from the ratio of the fluxes of the highly variable Mg II h and k emission cores (chromospheric origin) to that of the weakly variable nearby wings (photospheric origin) as discussed in \cite{heath1986}. The index can be downloaded at the Bremen University site (\url{http://www.iup.uni-bremen.de/UVSAT/Datasets/mgii}) and it is commonly considered a good proxy of solar magnetic activity and UV and EUV irradiance since it supplies a good estimate of UV irradiance variability \citep{white1998,viereck1999,dewit2009}. However, as mentioned earlier, the index or proxy Mg II is a core-to-wing ratio of the Mg II line and it probes the high chromosphere of the Sun. Since it is an index, the overall information from the solar disk is compressed into a single number, losing the spatial information on the structures responsible for the emission. Currently no data is available of full disk spatially resolved emission from the Sun in the Mg II doublet. The Aditya-L1 mission, currently planned to be launched in late 2022, has the possibility to observe both h and k Mg II lines with its Solar Ultraviolet Imaging Telescope (SUIT) \citep{aditya}. The imaging of Mg II doublet has been largely exploited by the Interface Region Imaging Spectrograph (IRIS) in the last 8 years, mainly focusing on the rapid evolution during solar flares in a field of view of 120'' by side \citep[see e.g.][]{kerr2015}. A recent paper by \cite{Orozco2022} describe the CASPER solar mission proposed for the ESA "Fast" F mission, which include a 30 cm aperture telescope for spectropolarimetry imaging in the Mg II h and k lines. 

Regarding the UV images of the solar disk, we have to point out that the magnetic structures responsible for the local heating in the Sun's chromosphere, and which are ultimately responsible for solar UV variability (i.e., SSI variability), and in particular for the variability in the Mg II line, have a double astrophysical importance. On the one hand, they modulate the solar radiative flux, on the other they introduce a contribution (which astrophysicists call stellar noise) to the solar radial velocity signal due, in addition to the convective blueshift, also to the rotational imbalance due to the presence of these brightness inhomogeneities of magnetic origin. The UV Imager on-board SEE will make an important contribution in this direction because it will provide an additional band to deconvolve the solar radial speed signal \citep[see][]{Milbourne2019} from the contributions of solar magnetic activity.

The solar UV radiation has a significant impact on Earth’s atmosphere. Photons with a wavelength lower than 300 nm affect indeed many chemical cycles in the atmospheres, as Ozone’s cycle in Earth’s stratosphere. Additionally, the thermosphere, 80 to 600 km in altitude, is primarily heated by solar UV radiation \citep{Floyd2002} and, consequently, satellites or debris in LEO are significantly affected by the variable atmospheric drag modulated by solar radiation variability from Medium UV (MUV, as i.e. the 280 nm band observed by the SEE payload) to Extreme UV (EUV).
The solar UV photons dissociate molecules or ionize the neutral atmosphere and therefore modify the chemical composition of the atmospheres and its temperature distribution, ultimately creating a dynamics of the planetary atmospheres that is modulated by: i) by the rotation of the planets (the day-night cycle), ii) by the variations of irradiance linked to orbital factors, e.g., the Sun-planet distance, and finally iii) from solar variability in UV and higher energy ranges \citep[see, e.g.][]{snow2005,dewit2010,lovric2017,berrilli2020SoPh,woods2021}. While the first two effects are easily estimated from the laws of celestial mechanics, solar variability remains a major issue. Several models are used to reconstruct the measured SSI at different wavelength and in particular in the UV \citep[see, e.g.][]{SSIFligge1998,SSI2008A&A...486..311U,SSI2011JGRD..11620108F,SSI2019SciA....5.6548S}. Although some attempts have been done to forecast the TSI \citep[see, e.g.][]{SSI2022JSWSC} there is currently no predictive model of SSI that can be used to forecast the thermosphere conditions.

Moreover, the effects of solar activity on the stratosphere have a direct impact on the Earth atmospheric circulation \citep[e.g., ][]{gray2010}.
The mechanism involves the stratosphere absorption of the solar UV in the spectral band between 100 and 300 nm. At mid-latitudes this happens at a height between 10 and 50 km where UV photons have enough energy to alter the chemical reactions that create and destroy ozone or photo-dissociate molecular oxygen.
As demonstrated with impressive evidence by the recent loss of 40 out of 49 Starlink satellites of SpaceX due to a mild geomagnetic storm
triggered by a Space Weather event, forecasting the thermosphere environment is critical to space mission design, re-entry
operations and space surveillance applications. Indeed, most low-Earth-orbit (LEO) operational satellites fly in a limited zone
between 400 and 800 km within the thermosphere. The orbital decay rate of satellites depends on atmospheric drag, which is
directly affected by the variable solar input (UV radiation and secondary effects due to geomagnetic induced storms). The orbital
decay rate at 250 km of altitude is very significant, causing re-entry of a satellite in about 2 weeks. In a recent paper \cite{bigazzi2020} performed an analysis of three of the most used solar flux and geomagnetic proxies, i.e., F10.7 flux, Mg II UV proxy and geomagnetic index Ap, in relation to the time evolution of thermospheric density measured in situ, from 260 to 230 km altitude, using the GOCE thermospheric-density dataset.
In this study the Mg II has been proved to be a better proxy than F10.7 in
capturing the long-term trends of the solar input, providing a good representation of the thermospheric density signal at low thermospheric altitudes and underlining once more the importance of UV data in forecasting the thermosphere properties.

Moreover, ozone is a crucial component of Earth's atmosphere in terms of its energy balance: its spectral band centered at 9.6 $\mu$m is positioned nearby Earth's thermal emission peak, thus a variation in the ozone concentration modulates the quantity of infrared radiation emitted by the Earth’s surface. This effect is clearly visible, where higher ozone levels produce a warming of the lower stratosphere and cooling of the upper stratosphere, even though ozone is a minor component of Earth's atmosphere with a concentration of merely 0.02–0.1 ppmv. 
A secondary goal of the SEE mission is to acquire images of the Earth's atmosphere at 280 nm.
In this observing mode the main target is the dayglow in the middle ultraviolet (MUV) coming from upper layers of Earth atmosphere. The measured airglow at mid and high latitudes can track geomagnetic disturbances \citep[see e. g.,][]{Meier1991} and ionosphere response, directly measuring the geomagnetic effects of Space Weather. This would further improve the SEE science products in case of non-constant solar pointing of the UV imager.

At higher energies, the solar emission in the Soft-X-Rays (SXR), Hard-X-Rays (HXR) and gamma-rays is dominated by solar flares. 
The main scientific drivers for the SEE payload devoted to gamma and X-rays flux monitoring are the elusive scales of solar flares emission. The spatial scale involved in the magnetic reconnection driving solar flares is $\sim$1 m \citep[see e.g. ][]{Shibata2001} and therefore out of the observing capabilities of any current instrument. Nevertheless, some information on the physics of solar flare trigger could be encoded in the unexplored sub-seconds time scales.
Solar flares relate to the sudden release of energy stored in the solar magnetic field in the upper layers of solar atmosphere. The trigger mechanism is not completely understood and is commonly explained in terms of magnetic reconnection \citep{Giovanelli1946}. The energy release mainly manifests as a brightening along all the electromagnetic spectrum, with a duration of the order of minutes. We refer to the extensive review \cite{Benz2008} for a detailed explanation of the phenomenon from an observational point of view. The energy radiated as photons during a large flare event is mostly prominent in the SXR band, where can exceed the pre-flare emission level by a factor of 200 \citep{woods2004}. The same study underline the fact that most of the energy is nevertheless irradiated above the 200 nm (77\%). SXR emission is used to define solar flares classification (X, M, C, B) and deeply studied since 1976 by the NASA GOES missions.
Solar flares are the most powerful transient event happening in the Solar System. The magnetic energy is released under different forms: photons, accelerated particles, heat, waves and plasma motion. Summed up all together, an X flare can release more than 10$^{25}$ J of energy \citep{emslie2005}.
EUV and SXR emissions are interpreted from the heating of the plasma in the coronal loops at temperature from 1.5 MK to beyond 30 MK.
In the HXR the photon energy distribution follow a non-thermal shape, close to a power-law. The source in this case is the bremsstrahlung of the accelerated electrons colliding on the denser atmospheric layers below the flare site. Line emission in the gamma-ray is due to heavy nuclei emission excited by MeV protons \citep{chupp1973}.

Although SXR observations may suggest solar corona as the prominent site of flare energy release, solar flares involve different physical mechanism in a complex interplay between corona and chromosphere. SXR and EUV emissions characterize the preflare phase at the top of the magnetic loop while HXR emission sources are located at the magnetic footpoints during the impulsive phase \citep{Benz2008}.
However, observations to date have not provided a clear picture to distinguish among the different mechanisms involved in the plasma heating and particle acceleration.
The fast-time domain of solar flare spectroscopy in SXR and HXR could provide a new window to distinguish different acceleration mechanisms. In a recent paper \cite{knuth2020} performed an analysis of fast fluctuations in Fermi Gamma-ray Burst Monitor (Fermi
GBM) data from two M9.3 class solar flares occurred in 2011. They studied with subsecond resolution X-ray spikes and their quasiperiodic nature. Differences in the timing of these spikes across energies can reveal time of flight effects such as particle trapping or it can indicate a preferred acceleration mechanism. With just two events available no conclusive analysis could be made. To date fast-time domain analysis have been limited by the $\sim$ 4 s cadence of RHESSI. While STIX on board of Solar Orbiter currently provides $\sim$ 1 s scale observations \citep{piana2022}, observations at much higher temporal cadence would open completely new opportunities to test for acceleration models.
The SEE mission science objectives include the study of pre-flare and flare conditions in an extended range of energies (keV to MeV) at an unprecedent cadence (up to 10 kHz), to study the reconnection events that drives the flare events and to possibly distinguish particle acceleration mechanisms.
As the highest number of flares is reached during the maximum of the solar cycle, therefore, based on \citep{Penza2021} and solar cycle prediction panel (\url{https://www.swpc.noaa.gov/products/solar-cycle-progression}) the preferred time window of the SEE mission (1 yr nominal +1 yr extended) is the time span 2023-2028.

SEE mission will also have an impact on Space Weather assets, aimed to provide countermeasure to technology threats and impacts on astronauts health.
The SXR and HXR emissions of intense solar flares can change the Earth’s ionosphere, disturbing or disrupting radio transmissions and Global Navigation Satellite System (GNSS) signals. Solar flares are also associated with the launch of Coronal Mass Ejections (CMEs), which can ultimately impact Earth producing geomagnetic storms. SEE gamma and X-ray instrument will measure SXR flux in the same GOES bands that are used to define the standard flare classification. This will allow for cross-calibration with the lower X-ray energy bands of SEE, with the advantage of having higher cadence with respect to GOES. At the same time SEE mission will also measure the flux at higher energies, up to few MeV, that can expose astronauts on lunar and deep space extravehicular activities (EVAs) to dangerous acute biological doses \citep[see e.g.,][]{smith2007,narici2018}.
SEE will address key questions concerning the physics of the Sun and the Earth's upper atmosphere, namely:
\begin{enumerate}
    \item What determines the solar UV budget in a very variable spectral region of primary interest for Earth's upper atmosphere (thermosphere, ionosphere, stratosphere)?
    \item What are the solar active regions that contribute most to the UV signal at 280 nm?
    \item What are the high-rate temporal characteristics of the Gamma and X-ray signal in a pre-flaring condition or during the flare?
\end{enumerate}

Although SEE is mainly a scientific mission, aimed at improving the understanding of the behavior of the Sun in a spectral region of strong astrophysical and geophysical importance (i.e. the Mg II line and high-energy radiation), the mission has aspects of a technological demonstrator, using SiPM detectors for the detection of the Gamma and X-ray solar signals, and with applicative and service implications, since the data produced will complement useful data in the Space Weather context.
In such scenario SEE mission will provide exceptional multiwavelength data of Space Weather events during solar cycle 25, complementing data from:
\begin{itemize}
    \item in-orbit observatories and experiments such as SDO \citep{SDO}, GOES \citep{GOES}, IRIS \citep{IRIS}, HINODE \citep{HINODE}, MAXI \citep{MAXI}, AMS-02 \citep{AMS,ams2021}, Mini-EUSO \citep{Mini-EUSO};
    \item deep space missions as Solar Orbiter \citep{SO}, Parker Solar Probe \citep{PSP} and Bepi Colombo \citep{bepicolombo};
    \item ground based networks and synoptic observatories such as GONG \citep{gong}, Kanzelh{\"o}he Observatory \citep{graz}, MOTH \citep{moth} and TSST \citep{TSST}.
\end{itemize}
The SEE mission has been proposed in the “Future missions for Cubesat” call of the Italian Space Agency (ASI) in 2020, and it has passed the technical evaluation phase in 2021. ASI funded the SEE mission for phases A/B in July 2022.
\begin{figure}[ht]
    \centering
    \includegraphics[scale=0.85]{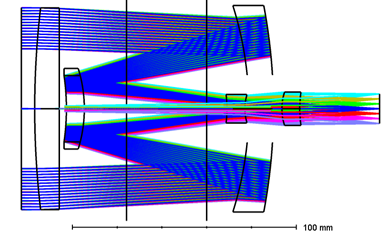}
    \caption{Preliminary optical layout of the SEE UV imager. The optical unit consists of 2 spherical mirrors and 3 spherical lenses, with a 90 mm aperture. 
    }
    \label{UVpayload}
\end{figure}
\section{UV imager}
\begin{table*}
    \centering
    \begin{tabular}{ccccccccc}
        \hline
        FOV & Bandpass & F/N & Focal length & Distortion & Obscuration & Vignetting & Barrel length & Max diameter\\
        \hline
        1.1 $^{\circ}$ & 2 nm & 7.6 & 690 mm & $<$1\% & $<$35\% & Negligible & 160 mm & 90 mm \\
        \hline
    \end{tabular}
    \caption{Technical characteristics of the telescope.}
    \label{tab_UV}
\end{table*}
The imager proposed for this mission allows to map the radiative emission of the Sun at 280.0 nm, with a passband of 2 nm, returning information on the solar regions responsible for the emission.
The field of view (FOV) is 1.1 $^\circ$, allowing full disk images compatible with the SEE platform pointing capabilities. The pixel scale is $\sim$2 arcsec/pixel.
These maps can be used, together with magnetograms and high resolution images of other missions (eg, SDO, IRIS, Solar Orbiter, ...) or with observations from ground-based synoptic telescopes in order to better understand the physical processes underlying the UV emission of the Sun and to create empirical models for the reconstruction of the spectral variations of the Sun.
The telescope is composed of 2 spherical mirrors and 3 spherical lenses. The compact design allows to fit the UV imager payload in less than 2U volume. A preliminary layout by Optec is shown in Fig. \ref{UVpayload}.
As reported in the table \ref{tab_UV}, the distortion is less than 1\% and the field curvature (i.e. the difference in best focus position from optical axis and marginal FOV) is less than the depth of focus of the instrument, causing negligible variation in terms of performances.
The vignetting is entirely caused by the central obscuration to the secondary mirror. 
In order to avoid loss in transmission due to the radiative environment, the selected material will be radiation hardened and selected to maximize transmission in UV range. In addition, in order to avoid thermal issues, two solutions will be implemented: focus mechanism (already tested and working in space); athermalization through fine selection of material. Transversal thermal gradient, nevertheless, can be compensated only by entire satellite athermalization. Another aspect that needs to be controlled is the straylight due to surface microroughness. Since the scattering is more sensitive in the UV wavelength range, surface microroughness will be accurately investigated.
The SEE science team will define the scientific requirements and the related technical requirements for the detailed design of the SEE UV imager based also on the heritage of the design of the ISODY payload \citep{isody} which included tunable interferometric filters \citep{fpi} proposed for the ADAHELI solar mission \citep{adaheli,adaheli2015}.

\section{Gamma and X-Ray instrument}
\begin{figure*}
    \centering
    \includegraphics[scale=0.7]{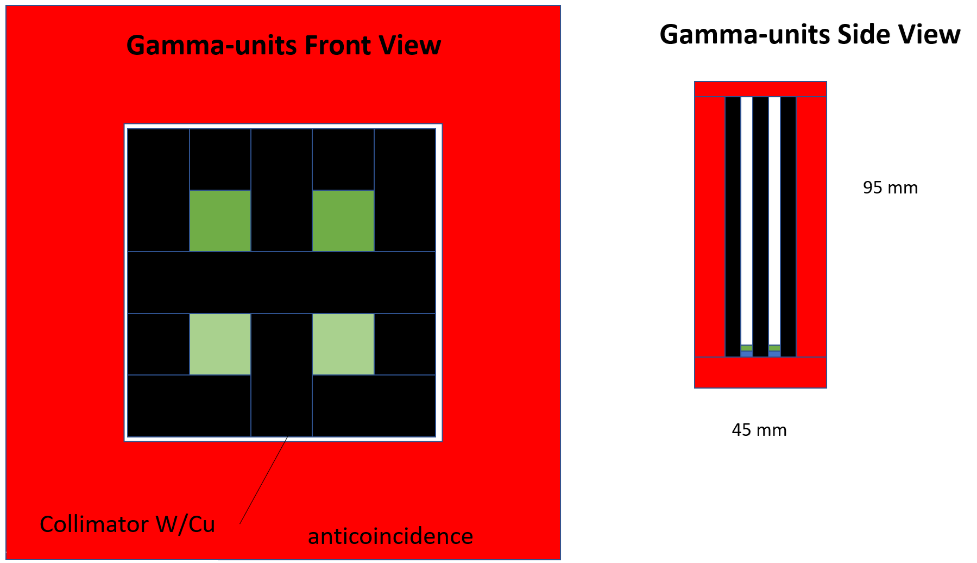}
    \includegraphics[scale=0.7]{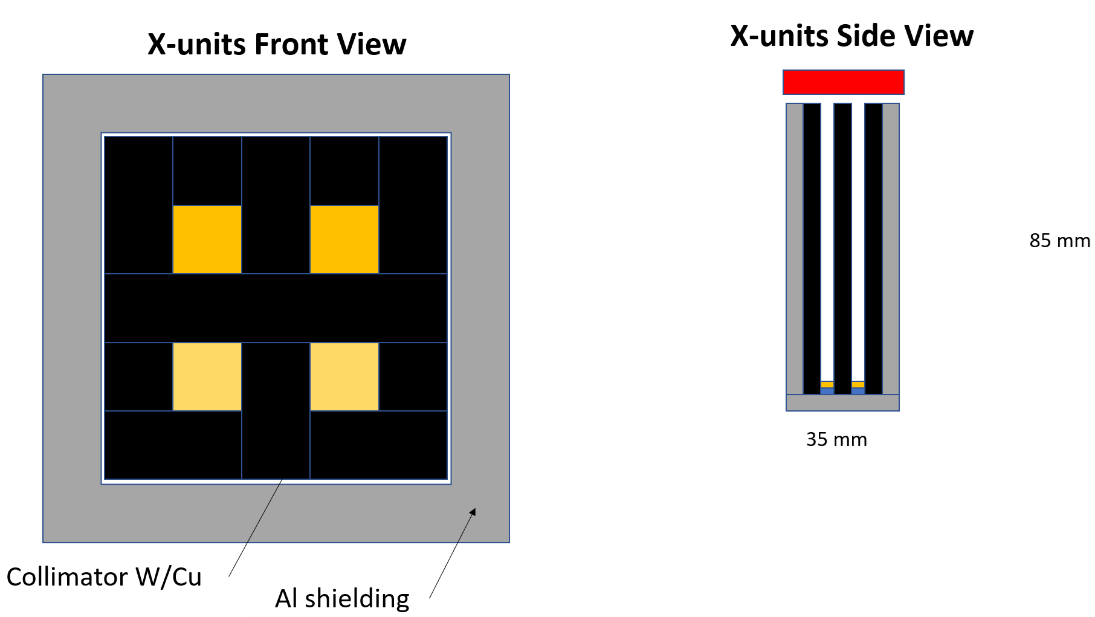}
    \caption{Preliminary layout of the SEE gamma (left) and X-ray (right) instrument.The SiPM/SP detectors are shown in green and yellow respectively. Anticoincidence crystals are shown in red, collimators are in black, aluminium shielding is in grey.
    }
    \label{x_gamma}
\end{figure*}
The high-energy payload is divided in two units: the gamma and the X-ray instrument. Each instrument is based on four (for redundancy and wider energy range) SiPM/Scintillator detectors. The readout uses SiPM (Silicon Photomultiplier, also known as MPPC) detectors which read the signal coming from the gamma ray conversion in the scintillator (CsI or Lyso). Two to four SiPMs read the same scintillator to improve the uniformity of readout and as additional redundancy, surrounded by an anticoincidence to reject charged particles.
The SiPM-based detectors will continuously acquire at a cadence of 10 kHz and in trigger mode, so to reveal the signal at the highest cadence.
In particular, the SiPM detector will measure the cascade of particles produced by each incoming  solar energetic photon that will be produced in the scintillator. The signal produced in this way is proportional to the energy of the solar photon, thus providing the possibility to tag each photon in time (with a time resolution of 0.1 ms or better) and energy.
It is therefore able to distinguish single bunch of photons (even single-photon events) at 10 kHz without losing the possibility to reveal all the photons available from flares of lower classes, for which the actual cadence of the observation will be lower.
Pre-amplification, readout and data acquisition are performed by a devoted board. The X-rays system is based on four SiPM/SP/Scintillator detectors. 
Both systems require a collimator (realized in Tungsten).
The whole system interfaces with the Payload computer for data compression and storage. The system is the evolution of similar ones already flown on previous/current missions on the International Space Station (ISS) in 2002 (Lazio-Sirad, employing for the first time SiPMs in space; as part of the Eneide mission, see \citealt{SiPM}) and subsequently in 2019 (Mini-EUSO as part of the Beyond Mission, employing multi-pixel SiPM arrays, see \citealt{Mini-EUSO}).
The detectors are arranged in a two-by-two grid as described in Fig. \ref{x_gamma}. The system can be divided in the following major components:
\begin{enumerate}
    \item Detector: It employs Hamamatsu MPPC (SiPM) for readout. A number of two to four MPPC will be used to ensure uniformity of readout of the scintillator and reliability in case of failure of one of the SiPMs. MPPCs are now currently used in a number of applications: the proponent group has a long-standing experience in using these detectors, since the first prototypes obtained from MepHI/Dolgoshein (the inventor of the technology) and in placing them in space: the already mentioned Lazio-Sirad mission and more recently as ancillary detectors on Mini-EUSO focal surface. 
    \item Scintillator. It is planned to use CsI or Lyso for their optimal gamma ray detection capabilities (e.g. density, ease of use, robustness…).  The optimal scintillator material(s) will be selected based on Montecarlo simulations and laboratory tests.
    \item Collimator. A tungsten collimator will be used to allow detection only of gamma rays coming from the Sun’s direction.
    \item Anticoincidence. A plastic anticoincidence system, surrounding the detector, will be used to reject the signal coming from charged particles (mostly electrons and protons) of cosmic radiation. The anticoincidence system will also be read by SiPMs, with the corresponding digital signal used as a veto to the trigger. 
    \item Frontend/readout electronics. Front-end signal readout and shaping will be performed by a devoted board which will amplify the signal, convert it to digital value, perform trigger duties. The board trigger will ensure a SNR of at least 10 for the lowest energy photon in each band. For example, the instrument in the 1-3 MeV range will have a trigger at 100 keV to detect the 1 MeV incoming photon with a SNR of 10. The board will also be devoted to temperature control and conditioning of the SiPMs, compensating for temperature changes in order to keep the gain of the detector stable.
\end{enumerate}

A baseline of 4 mm side sensor and 80 mm long collimator has been considered, providing an acceptance angle of 6$^{\circ}$. The current configuration consists of 8 scientific channels and 6 anticoincidence channels. Expected photon flux for a flare $>$X5 at different energies is available in table \ref{tab_x}. 
\cite{Aschwanden2012} provided a detailed analysis of the solar flare rates at different energies over 37 years (1975-2011). In this work, the total number of flares that have been detected larger than a given magnitude (X/M/C) decreases approximately by a factor of 10 for each order of magnitude. If we avoid a phase of solar minimum and considering the solar flare rates detected shown in Figure 3 in \cite{Aschwanden2012}, we can expect, over the one year nominal life-time of SEE, $\sim$ 100 flares $>$M1 and $\sim$ 10 flares $>$X1. We therefore include in the last two column of table \ref{tab_x} an estimate of the counts for M1 and X1 flares.
SEE science for the Gamma and X-ray instrument is driven by two main goals:
one the one hand the possibility to observe with sub-second cadence the more rare high energy flares ($>$X5); on the other hand by measuring extended energy range solar flare spectra for the more frequent $>$M1 flares.
Indeed, the detection of multiple photons at sub second resolution for the higher energy channels will be achieved only for very high energy flares, nevertheless the time tag of even a single photon compared with the time and energy distribution in the X-ray channels will also provide valuable information on the flare acceleration process.
The estimated volume and mass for the X-rays units, the gamma-rays units and the frontend/readout electronics are approximately 1 U and 3 kg respectively. The power consumption of the X/gamma payload is 15 W. 
\begin{table*}
    \centering
    \begin{tabular}{lccccccc}
        \hline
        Band & Source & Flux ($>$X5) & Sensor & Counts in the &  Channel & Channel & Channel \\
        & & (counts keV$^{-1}$ & bandwidth & band ($>$X5) & counts ($>$X5) & counts ($>$X1) & counts ($>$M1)\\
        &  & cm$^{-2}$ s$^{-1}$) & (keV) & (counts cm$^{-2}$ s$^{-1}$) & (counts s$^{-1}$) & (counts s$^{-1}$) & (counts s$^{-1}$) \\
        
        \hline
        & & & & & & & \\
        3 keV & thermal & 10$^7$ & 10 & 10$^8$ & 10$^7$ & 2x10$^6$ & 2x10$^5$ \\
        (GOES & & & (1.5-12keV) & & & & \\
        XRS-B) & & & & & & & \\
         & & & & & & & \\
        6 keV & thermal & 10$^6$ & 20 & 2x10$^7$ & 10$^6$ & 2x10$^5$ & 2x10$^4$ \\
        (GOES  & & & (3-25keV) & & & & \\
        XRS-A) & & & & & & & \\
        & & & & & & & \\
        100-200 keV & non-thermal & 10 & 10 & 10$^2$ & 10 & 2 & 2x10$^{-1}$ \\
        & & & & & & & \\
        1-3 MeV & continuum +  & 10$^{-2}$ & 1000 & 10 & 1 & 2x10$^{-1}$ & 2x10$^{-2}$\\
        & lines & & & & & & \\
        \hline
    \end{tabular}
    \caption{Expected photon flux for a flare $>$X5 at different energies, based on \citet{Lin2002}. The last three columns show the expected number of counts on each sensor channel respectively for an X5, X1 and M1 -class flares. They have been computed scaling the X5-class flare. The computations are supported also by other flare spectra, see e.g. \cite{Grigis2004}.}
    \label{tab_x}
\end{table*}
\section{Mission profile and CubeSat platform}
During the Nominal Mission, SEE shall constantly point at the Sun, except during manoeuvres, eclipses or contingencies. The UV Imager will observe the Sun at regular intervals (baseline every 6h) to obtain synoptic images of the solar disk while the Gamma and X-ray detectors will acquire the integrated solar signal continuously.
This high energy signal will be used as a trigger, activated by intense solar flares, to activate a high-cadence mode of acquisition of the Imager.
The satellite radial velocity in the Sunward direction shall not exceed +/-4 km/s during the orbit, keeping the orbital shift Doppler within 4x10$^{-3}$ nm. Given the 2 nm passband of the filter centered at 280.0 nm, this ensures to always include the Mg II h (276.56 nm) and k (280.27) nm lines, having an intensity contamination from the wings of the line $<$ 0.1\%.
The accuracy in pointing at the solar disk (given its position w.r.t. the center of the Sun’s disk) must be a fraction of the field of view of the Imager, say $<$ 1.1$^{\circ}$ (to get the solar disk within the field of view of the high resolution telescope).
In order to fulfill science requirements (i.e., synoptic images of the Sun at 280 nm and quasi-continuous Gamma and X-ray monitoring), two observation modes are preliminary foreseen.
\begin{enumerate}[a)]
    \item NOMINAL (NO-FLARE) MODE. Four Mg II (280 nm) full disk images per day, every 6 hours. The gamma and x-ray instrument will acquire continuously with single-photon detection capabilities at up to 10 kHz sampling (advanced goal 1 MHz; GOES provides 1 Hz fluxes), except during manoeuvres, eclipses or contingencies. 
    \item FAST (FLARE) MODE (TBC). Trigger from X-ray channel to the UV imager. Mg II (280 nm) full disk images every 60 seconds. Burst of continuous sampling up to 10 kHz (advanced goal 1 MHz; GOES provides 1 Hz fluxes) of the gamma and x-ray instrument (when pointed).
\end{enumerate}

\begin{figure}[ht]
    \centering
    \includegraphics[scale=0.95]{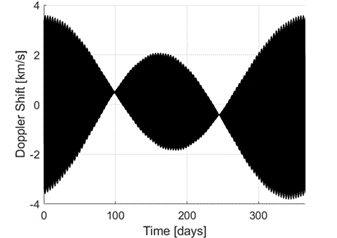}
    \caption{Doppler shift for the 550 km circular SSO.
    }
    \label{orbit}
\end{figure}

The orbit chosen for the SEE is a circular Sun Synchronous Orbit (SSO). The SSO is a retrograde orbit (inclination greater than 90 degrees) where the satellite passes over any given point of the planet's surface at the same local mean solar time and the orbital plane normal is always directed towards the Sun direction. The SSO allows SEE to have low eclipse times, which are useful for the observations and the solar panels. For this analysis, the eccentricity has been set equal to 0 to obtain a circular orbit and the altitude actually represents the design variable (the inclination is retrieved accordingly). In particular, several altitudes have been considered ranging from 550 km to 750 km, although they present different advantages. For example, a lower altitude SSO requires lower (or even totally null) propellant consumption for de-orbiting and lower power for data transmission. On the other hand, a higher altitude SSO allows a lower eclipse time and higher visibility time with a ground station. However, all the altitudes can potentially be suitable to achieve the SEE scientific objectives. 
For the analysis presented in this paper, a 550 km circular SSO is chosen as an example. For the selected orbit, the trend of the satellite radial velocity is shown in Fig. \ref{orbit}.
Considering that the SEE payload can be hosted in 4 CubeSat Units (4U) with minor modifications on the current payload configuration, the Argotec HAWK platform can be exploited in its 12U CubeSat Configuration. The HAWK platform supports the satellite’s operations and functionalities with the following subsystems:
\begin{itemize}
    \item Structure (SS), which is in charge of providing the physical support for the required hardware and of withstanding launch, deployment and operational mechanical loads; 
    \item Attitude Determination and Control (ADCS), whose aim is to determinate and control attitude, considering where the satellite should point and thus orient it;
    \item On-Board Computer \& Data Handling (OBC\&DH), that provides communication between all the subsystems, to guarantee their right interaction, and to perform the desired satellite’s operations;
    \item Electrical Power (EPS), which is in charge of providing, managing and storing the required electrical power. Its aim is also to convert and distribute the electrical power coming from the SPA to the Battery;
    \item Telemetry Tracking \& Command (TT\&C), that allows the satellite to transmit data to the ground, including scientific acquisitions. It guarantees also the reception of commands sent from ground;
    \item Thermal Control (TCS), which keeps all the subsystems’ components in their prescribed temperature range.
\end{itemize}
\begin{figure*}[ht]
    \centering
    \includegraphics[scale=0.7]{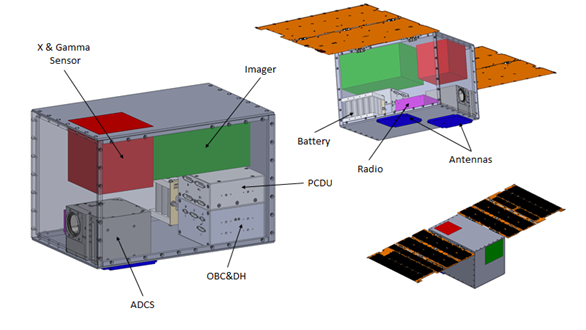}
    \caption{Preliminary subsystems accommodation in the SEE cubesat.
    }
    \label{cubesat}
\end{figure*}
A preliminary analysis was carried out in order to define the baseline subsystems’ configuration for the CubeSat, shown in Fig. \ref{cubesat}. The configuration design was forced by the following constraints:
constant Sun-pointing may degrade, and most likely damage, the CMOS. Ageing will be assessed after detailed analysis of the rejection and narrow band filter. A possible solution is the use of a shutter in the UV payload.
The X \& Gamma sensor shall point the Sun for long periods of time.
According to these constraints, the preliminary subsystem accomodation is reported in Fig. \ref{cubesat}.

The satellite will be in Sun Pointing mode most of its lifetime, with the exception of
communication windows. 
A preliminary mass and power budget shows that the CubeSat complies with the standard 12U CubeSat requirement (mass requirement: 24 kg) considering the proper margins at subsystem and system level.
Moving forward to a preliminary data and link budget, the HAWK platform embarks an OBC\&DH able to store up to 16GB of data. The imager will take no more than 4 pictures everyday, with a data volume of 24MB/day if no compression method is used. The X and Gamma Ray sensor will acquire data constantly, creating a data volume of 26MB/day. With a total of 100MB/day (with a taken margin of 100\%), the OBC\&DH can store data for 160 days.

\section{Conclusions}

SEE mission will be a pathfinder: a low-cost solar mission based on CubeSat approach, with an expected high scientific output. SEE will investigate the cornerstone Mg II region of the solar spectrum, with imaging capabilities. SEE UV imager has the goal to unveil the roots of solar UV variations and its connection with magnetic structures in the solar atmosphere over different time scales. It will provide a new perspective on solar UV emission in combination with data from present (Solar Orbiter, SDO, GOES) and future (Vigil, Solar-C, Aditya-L1) large solar space projects.
Furthermore, SEE will explore for the first time the solar SXR and HXR spectrum at fast cadence (up to 10 kHz and with single-photon detection capabilities), searching for pre-flare and trigger signals and trying to distinguish the still not completely understood particle acceleration mechanism during flare events.
The expected user base is potentially very large, as the data provided by the SEE mission is highly relevant for many scientific communities: astrophysics, geophysics, space weather. The Italian Space Agency may also have a specific interest in the data produced. In fact, ASI is working on the creation of an Italian Space Weather Infrastructure: the ASPIS project \citep{plainaki2020}.
SEE will certainly represent a test-bed for new technologies which are to be employed in future missions, such as SiPM for SXR and HXR spectroscopy at high cadence, compact UV telescopes for CubeSats platforms, new materials and coating for mirrors, low-cost memories.
The specific expertise on CubeSats missions of the industrial partners, ARGOTEC s.r.l, OPTEC S.p.A., involved in the ArgoMoon and LICIAcube missions, and the expertise of NEXT Ingegneria dei Sistemi S.p.A. in the ground segment and operations definition, has perfectly supported the Science Team in defining the set of requirements compatible with the limited budget, yet leaving untouched SEE’s innovative capabilities.

\section{Acknowledgments}

This research is partially supported by the Italian MIUR-PRIN
\emph{Circumterrestrial Environment: Impact of Sun-Earth Interaction\/}
grant 2017APKP7T and by financial support from the European Union’s
Horizon 2020 research and innovation program under grant agreement No.
824135 (SOLARNET).
M.C., P.G. and R.R. are PhD students of the PhD course in Astronomy, Astrophysics and Space Science, a joint research program between the University of Rome “Tor Vergata”, the Sapienza University of Rome and the National Institute of Astrophysics (INAF).
G.F. and R.M. are PhD students of the SWATNet project.
SWATNet has received funding from the European Union’s Horizon 2020 research and innovation programme under the Marie Sklodowska-Curie Grant Agreement No 955620.

\bibliographystyle{model5-names}
\biboptions{authoryear}
\bibliography{refs}

\begin{thebibliography}{61}
\expandafter\ifx\csname natexlab\endcsname\relax\def\natexlab#1{#1}\fi
\providecommand{\url}[1]{\texttt{#1}}
\providecommand{\href}[2]{#2}
\providecommand{\path}[1]{#1}
\providecommand{\DOIprefix}{doi:}
\providecommand{\ArXivprefix}{arXiv:}
\providecommand{\URLprefix}{URL: }
\providecommand{\Pubmedprefix}{pmid:}
\providecommand{\doi}[1]{\href{http://dx.doi.org/#1}{\path{#1}}}
\providecommand{\Pubmed}[1]{\href{pmid:#1}{\path{#1}}}
\providecommand{\bibinfo}[2]{#2}
\ifx\xfnm\relax \def\xfnm[#1]{\unskip,\space#1}\fi
\bibitem[{{Aguilar} et~al.(2013){Aguilar}, {Alberti}, {Alpat}, {Alvino},
  {Ambrosi}, {Andeen}, {Anderhub}, {Arruda}, {Azzarello}, {Bachlechner},
  {Barao}, {Baret}, {Barrau}, {Barrin}, {Bartoloni}, {Basara}, {Basili},
  {Batalha}, {Bates}, {Battiston}, {Bazo}, {Becker}, {Becker}, {Behlmann},
  {Beischer}, {Berdugo}, {Berges}, {Bertucci}, {Bigongiari}, {Biland}, {Bindi},
  {Bizzaglia}, {Boella}, {de Boer}, {Bollweg}, {Bolmont}, {Borgia}, {Borsini},
  {Boschini}, {Boudoul}, {Bourquin}, {Brun}, {Bu{\'e}nerd}, {Burger}, {Burger},
  {Cadoux}, {Cai}, {Capell}, {Casadei}, {Casaus}, {Cascioli}, {Castellini},
  {Cernuda}, {Cervelli}, {Chae}, {Chang}, {Chen}, {Chen}, {Chen}, {Cheng},
  {Chen}, {Cheng}, {Chernoplyiokov}, {Chikanian}, {Choumilov}, {Choutko},
  {Chung}, {Clark}, {Clavero}, {Coignet}, {Commichau}, {Consolandi}, {Contin},
  {Corti}, {Costado Dios}, {Coste}, {Crespo}, {Cui}, {Dai}, {Delgado}, {Della
  Torre}, {Demirkoz}, {Dennett}, {Derome}, {Di Falco}, {Diao}, {Diago},
  {Djambazov}, {D{\'\i}az}, {von Doetinchem}, {Du}, {Dubois}, {Duperay},
  {Duranti}, {D'Urso}, {Egorov}, {Eline}, {Eppling}, {Eronen} et~al.}]{AMS}
\bibinfo{author}{{Aguilar}, M.}, \bibinfo{author}{{Alberti}, G.},
  \bibinfo{author}{{Alpat}, B.}, \bibinfo{author}{{Alvino}, A.},
  \bibinfo{author}{{Ambrosi}, G.}, \bibinfo{author}{{Andeen}, K.},
  \bibinfo{author}{{Anderhub}, H.}, \bibinfo{author}{{Arruda}, L.},
  \bibinfo{author}{{Azzarello}, P.}, \bibinfo{author}{{Bachlechner}, A.},
  \bibinfo{author}{{Barao}, F.}, \bibinfo{author}{{Baret}, B.},
  \bibinfo{author}{{Barrau}, A.}, \bibinfo{author}{{Barrin}, L.},
  \bibinfo{author}{{Bartoloni}, A.}, \bibinfo{author}{{Basara}, L.},
  \bibinfo{author}{{Basili}, A.}, \bibinfo{author}{{Batalha}, L.},
  \bibinfo{author}{{Bates}, J.}, \bibinfo{author}{{Battiston}, R.},
  \bibinfo{author}{{Bazo}, J.}, \bibinfo{author}{{Becker}, R.},
  \bibinfo{author}{{Becker}, U.}, \bibinfo{author}{{Behlmann}, M.},
  \bibinfo{author}{{Beischer}, B.}, \bibinfo{author}{{Berdugo}, J.},
  \bibinfo{author}{{Berges}, P.}, \bibinfo{author}{{Bertucci}, B.},
  \bibinfo{author}{{Bigongiari}, G.}, \bibinfo{author}{{Biland}, A.},
  \bibinfo{author}{{Bindi}, V.}, \bibinfo{author}{{Bizzaglia}, S.},
  \bibinfo{author}{{Boella}, G.}, \bibinfo{author}{{de Boer}, W.},
  \bibinfo{author}{{Bollweg}, K.}, \bibinfo{author}{{Bolmont}, J.},
  \bibinfo{author}{{Borgia}, B.}, \bibinfo{author}{{Borsini}, S.},
  \bibinfo{author}{{Boschini}, M.~J.}, \bibinfo{author}{{Boudoul}, G.},
  \bibinfo{author}{{Bourquin}, M.}, \bibinfo{author}{{Brun}, P.},
  \bibinfo{author}{{Bu{\'e}nerd}, M.}, \bibinfo{author}{{Burger}, J.},
  \bibinfo{author}{{Burger}, W.}, \bibinfo{author}{{Cadoux}, F.},
  \bibinfo{author}{{Cai}, X.~D.}, \bibinfo{author}{{Capell}, M.},
  \bibinfo{author}{{Casadei}, D.}, \bibinfo{author}{{Casaus}, J.},
  \bibinfo{author}{{Cascioli}, V.}, \bibinfo{author}{{Castellini}, G.},
  \bibinfo{author}{{Cernuda}, I.}, \bibinfo{author}{{Cervelli}, F.},
  \bibinfo{author}{{Chae}, M.~J.}, \bibinfo{author}{{Chang}, Y.~H.},
  \bibinfo{author}{{Chen}, A.~I.}, \bibinfo{author}{{Chen}, C.~R.},
  \bibinfo{author}{{Chen}, H.}, \bibinfo{author}{{Cheng}, G.~M.},
  \bibinfo{author}{{Chen}, H.~S.}, \bibinfo{author}{{Cheng}, L.},
  \bibinfo{author}{{Chernoplyiokov}, N.}, \bibinfo{author}{{Chikanian}, A.},
  \bibinfo{author}{{Choumilov}, E.}, \bibinfo{author}{{Choutko}, V.},
  \bibinfo{author}{{Chung}, C.~H.}, \bibinfo{author}{{Clark}, C.},
  \bibinfo{author}{{Clavero}, R.}, \bibinfo{author}{{Coignet}, G.},
  \bibinfo{author}{{Commichau}, V.}, \bibinfo{author}{{Consolandi}, C.},
  \bibinfo{author}{{Contin}, A.}, \bibinfo{author}{{Corti}, C.},
  \bibinfo{author}{{Costado Dios}, M.~T.}, \bibinfo{author}{{Coste}, B.},
  \bibinfo{author}{{Crespo}, D.}, \bibinfo{author}{{Cui}, Z.},
  \bibinfo{author}{{Dai}, M.}, \bibinfo{author}{{Delgado}, C.},
  \bibinfo{author}{{Della Torre}, S.}, \bibinfo{author}{{Demirkoz}, B.},
  \bibinfo{author}{{Dennett}, P.}, \bibinfo{author}{{Derome}, L.},
  \bibinfo{author}{{Di Falco}, S.}, \bibinfo{author}{{Diao}, X.~H.},
  \bibinfo{author}{{Diago}, A.}, \bibinfo{author}{{Djambazov}, L.},
  \bibinfo{author}{{D{\'\i}az}, C.}, \bibinfo{author}{{von Doetinchem}, P.},
  \bibinfo{author}{{Du}, W.~J.}, \bibinfo{author}{{Dubois}, J.~M.},
  \bibinfo{author}{{Duperay}, R.}, \bibinfo{author}{{Duranti}, M.},
  \bibinfo{author}{{D'Urso}, D.}, \bibinfo{author}{{Egorov}, A.},
  \bibinfo{author}{{Eline}, A.}, \bibinfo{author}{{Eppling}, F.~J.},
  \bibinfo{author}{{Eronen}, T.} et~al. (\bibinfo{year}{2013}).
\newblock \bibinfo{title}{{First Result from the Alpha Magnetic Spectrometer on
  the International Space Station: Precision Measurement of the Positron
  Fraction in Primary Cosmic Rays of 0.5-350 GeV}}.
\newblock {\it \bibinfo{journal}{\prl}\/},  {\it
  \bibinfo{volume}{110}\/}\bibinfo{issue}{(14)}, \bibinfo{pages}{141102}.
  \DOIprefix\doi{10.1103/PhysRevLett.110.141102}.
\bibitem[{{Aguilar} et~al.(2021){Aguilar}, {Ali Cavasonza}, {Ambrosi},
  {Arruda}, {Attig}, {Barao}, {Barrin}, {Bartoloni}, {Ba{\c{s}}e{\u{g}}mez-du
  Pree}, {Bates}, {Battiston}, {Behlmann}, {Beischer}, {Berdugo}, {Bertucci},
  {Bindi}, {de Boer}, {Bollweg}, {Borgia}, {Boschini}, {Bourquin}, {Bueno},
  {Burger}, {Burger}, {Burmeister}, {Cai}, {Capell}, {Casaus}, {Castellini},
  {Cervelli}, {Chang}, {Chen}, {Chen}, {Chen}, {Cheng}, {Chou}, {Chouridou},
  {Choutko}, {Chung}, {Clark}, {Coignet}, {Consolandi}, {Contin}, {Corti},
  {Cui}, {Dadzie}, {Dai}, {Delgado}, {Della Torre}, {Demirk{\"o}z}, {Derome},
  {Di Falco}, {Di Felice}, {D{\'\i}az}, {Dimiccoli}, {von Doetinchem}, {Dong},
  {Donnini}, {Duranti}, {Egorov}, {Eline}, {Feng}, {Fiandrini}, {Fisher},
  {Formato}, {Freeman}, {Galaktionov}, {G{\'a}mez}, {Garc{\'\i}a-L{\'o}pez},
  {Gargiulo}, {Gast}, {Gebauer}, {Gervasi}, {Giovacchini}, {G{\'o}mez-Coral},
  {Gong}, {Goy}, {Grabski}, {Grandi}, {Graziani}, {Guo}, {Haino}, {Han},
  {Hashmani}, {He}, {Heber}, {Hsieh}, {Hu}, {Huang}, {Hungerford}, {Incagli},
  {Jang}, {Jia}, {Jinchi}, {Kanishev}, {Khiali}, {Kim}, {Kirn}, {Konyushikhin}
  et~al.}]{ams2021}
\bibinfo{author}{{Aguilar}, M.}, \bibinfo{author}{{Ali Cavasonza}, L.},
  \bibinfo{author}{{Ambrosi}, G.}, \bibinfo{author}{{Arruda}, L.},
  \bibinfo{author}{{Attig}, N.}, \bibinfo{author}{{Barao}, F.},
  \bibinfo{author}{{Barrin}, L.}, \bibinfo{author}{{Bartoloni}, A.},
  \bibinfo{author}{{Ba{\c{s}}e{\u{g}}mez-du Pree}, S.},
  \bibinfo{author}{{Bates}, J.}, \bibinfo{author}{{Battiston}, R.},
  \bibinfo{author}{{Behlmann}, M.}, \bibinfo{author}{{Beischer}, B.},
  \bibinfo{author}{{Berdugo}, J.}, \bibinfo{author}{{Bertucci}, B.},
  \bibinfo{author}{{Bindi}, V.}, \bibinfo{author}{{de Boer}, W.},
  \bibinfo{author}{{Bollweg}, K.}, \bibinfo{author}{{Borgia}, B.},
  \bibinfo{author}{{Boschini}, M.~J.}, \bibinfo{author}{{Bourquin}, M.},
  \bibinfo{author}{{Bueno}, E.~F.}, \bibinfo{author}{{Burger}, J.},
  \bibinfo{author}{{Burger}, W.~J.}, \bibinfo{author}{{Burmeister}, S.},
  \bibinfo{author}{{Cai}, X.~D.}, \bibinfo{author}{{Capell}, M.},
  \bibinfo{author}{{Casaus}, J.}, \bibinfo{author}{{Castellini}, G.},
  \bibinfo{author}{{Cervelli}, F.}, \bibinfo{author}{{Chang}, Y.~H.},
  \bibinfo{author}{{Chen}, G.~M.}, \bibinfo{author}{{Chen}, H.~S.},
  \bibinfo{author}{{Chen}, Y.}, \bibinfo{author}{{Cheng}, L.},
  \bibinfo{author}{{Chou}, H.~Y.}, \bibinfo{author}{{Chouridou}, S.},
  \bibinfo{author}{{Choutko}, V.}, \bibinfo{author}{{Chung}, C.~H.},
  \bibinfo{author}{{Clark}, C.}, \bibinfo{author}{{Coignet}, G.},
  \bibinfo{author}{{Consolandi}, C.}, \bibinfo{author}{{Contin}, A.},
  \bibinfo{author}{{Corti}, C.}, \bibinfo{author}{{Cui}, Z.},
  \bibinfo{author}{{Dadzie}, K.}, \bibinfo{author}{{Dai}, Y.~M.},
  \bibinfo{author}{{Delgado}, C.}, \bibinfo{author}{{Della Torre}, S.},
  \bibinfo{author}{{Demirk{\"o}z}, M.~B.}, \bibinfo{author}{{Derome}, L.},
  \bibinfo{author}{{Di Falco}, S.}, \bibinfo{author}{{Di Felice}, V.},
  \bibinfo{author}{{D{\'\i}az}, C.}, \bibinfo{author}{{Dimiccoli}, F.},
  \bibinfo{author}{{von Doetinchem}, P.}, \bibinfo{author}{{Dong}, F.},
  \bibinfo{author}{{Donnini}, F.}, \bibinfo{author}{{Duranti}, M.},
  \bibinfo{author}{{Egorov}, A.}, \bibinfo{author}{{Eline}, A.},
  \bibinfo{author}{{Feng}, J.}, \bibinfo{author}{{Fiandrini}, E.},
  \bibinfo{author}{{Fisher}, P.}, \bibinfo{author}{{Formato}, V.},
  \bibinfo{author}{{Freeman}, C.}, \bibinfo{author}{{Galaktionov}, Y.},
  \bibinfo{author}{{G{\'a}mez}, C.}, \bibinfo{author}{{Garc{\'\i}a-L{\'o}pez},
  R.~J.}, \bibinfo{author}{{Gargiulo}, C.}, \bibinfo{author}{{Gast}, H.},
  \bibinfo{author}{{Gebauer}, I.}, \bibinfo{author}{{Gervasi}, M.},
  \bibinfo{author}{{Giovacchini}, F.}, \bibinfo{author}{{G{\'o}mez-Coral},
  D.~M.}, \bibinfo{author}{{Gong}, J.}, \bibinfo{author}{{Goy}, C.},
  \bibinfo{author}{{Grabski}, V.}, \bibinfo{author}{{Grandi}, D.},
  \bibinfo{author}{{Graziani}, M.}, \bibinfo{author}{{Guo}, K.~H.},
  \bibinfo{author}{{Haino}, S.}, \bibinfo{author}{{Han}, K.~C.},
  \bibinfo{author}{{Hashmani}, R.~K.}, \bibinfo{author}{{He}, Z.~H.},
  \bibinfo{author}{{Heber}, B.}, \bibinfo{author}{{Hsieh}, T.~H.},
  \bibinfo{author}{{Hu}, J.~Y.}, \bibinfo{author}{{Huang}, Z.~C.},
  \bibinfo{author}{{Hungerford}, W.}, \bibinfo{author}{{Incagli}, M.},
  \bibinfo{author}{{Jang}, W.~Y.}, \bibinfo{author}{{Jia}, Y.},
  \bibinfo{author}{{Jinchi}, H.}, \bibinfo{author}{{Kanishev}, K.},
  \bibinfo{author}{{Khiali}, B.}, \bibinfo{author}{{Kim}, G.~N.},
  \bibinfo{author}{{Kirn}, T.}, \bibinfo{author}{{Konyushikhin}, M.} et~al.
  (\bibinfo{year}{2021}).
\newblock \bibinfo{title}{{The Alpha Magnetic Spectrometer (AMS) on the
  international space station: Part II - Results from the first seven years}}.
\newblock {\it \bibinfo{journal}{Physics Reports}\/},  {\it
  \bibinfo{volume}{894}\/}, \bibinfo{pages}{1--116}.
  \DOIprefix\doi{10.1016/j.physrep.2020.09.003}.
\bibitem[{{Aschwanden} \& {Freeland}(2012)}]{Aschwanden2012}
\bibinfo{author}{{Aschwanden}, M.~J.}, \& \bibinfo{author}{{Freeland}, S.~L.}
  (\bibinfo{year}{2012}).
\newblock \bibinfo{title}{{Automated Solar Flare Statistics in Soft X-Rays over
  37 Years of GOES Observations: The Invariance of Self-organized Criticality
  during Three Solar Cycles}}.
\newblock {\it \bibinfo{journal}{\apj}\/},  {\it
  \bibinfo{volume}{754}\/}\bibinfo{issue}{(2)}, \bibinfo{pages}{112}.
  \DOIprefix\doi{10.1088/0004-637X/754/2/112}.
  \href{http://arxiv.org/abs/1205.6712}{\tt arXiv:1205.6712}.
\bibitem[{{Bacholle} et~al.(2021){Bacholle}, {Barrillon}, {Battisti}, {Belov},
  {Bertaina}, {Bisconti}, {Blaksley}, {Blin-Bondil}, {Cafagna}, {Cambi{\`e}},
  {Capel}, {Casolino}, {Crisconio}, {Churilo}, {Cotto}, {de la Taille},
  {Djakonow}, {Ebisuzaki}, {Fenu}, {Franceschi}, {Fuglesang}, {Gorodetzky},
  {Haungs}, {Kajino}, {Kasuga}, {Khrenov}, {Klimov}, {Kochepasov}, {Kuznetsov},
  {Marcelli}, {Marsza{\l}}, {Mignone}, {Mascetti}, {Miyamoto}, {Murashov},
  {Napolitano}, {Olinto}, {Ohmori}, {Osteria}, {Panasyuk}, {Porfilio},
  {Poroshin}, {Parizot}, {Picozza}, {Piotrowski}, {Plebaniak},
  {Pr{\'e}v{\^o}t}, {Przybylak}, {Reali}, {Ricci}, {Sakaki}, {Shinozaki},
  {Szabelski}, {Takizawa}, {Turriziani}, {Tra{\"\i}che}, {Valentini}, {Wada},
  {Wiencke}, {Yashin} \& {Zuccaro-Marchi}}]{Mini-EUSO}
\bibinfo{author}{{Bacholle}, S.}, \bibinfo{author}{{Barrillon}, P.},
  \bibinfo{author}{{Battisti}, M.}, \bibinfo{author}{{Belov}, A.},
  \bibinfo{author}{{Bertaina}, M.}, \bibinfo{author}{{Bisconti}, F.},
  \bibinfo{author}{{Blaksley}, C.}, \bibinfo{author}{{Blin-Bondil}, S.},
  \bibinfo{author}{{Cafagna}, F.}, \bibinfo{author}{{Cambi{\`e}}, G.},
  \bibinfo{author}{{Capel}, F.}, \bibinfo{author}{{Casolino}, M.},
  \bibinfo{author}{{Crisconio}, M.}, \bibinfo{author}{{Churilo}, I.},
  \bibinfo{author}{{Cotto}, G.}, \bibinfo{author}{{de la Taille}, C.},
  \bibinfo{author}{{Djakonow}, A.}, \bibinfo{author}{{Ebisuzaki}, T.},
  \bibinfo{author}{{Fenu}, F.}, \bibinfo{author}{{Franceschi}, A.},
  \bibinfo{author}{{Fuglesang}, C.}, \bibinfo{author}{{Gorodetzky}, P.},
  \bibinfo{author}{{Haungs}, A.}, \bibinfo{author}{{Kajino}, F.},
  \bibinfo{author}{{Kasuga}, H.}, \bibinfo{author}{{Khrenov}, B.},
  \bibinfo{author}{{Klimov}, P.}, \bibinfo{author}{{Kochepasov}, S.},
  \bibinfo{author}{{Kuznetsov}, V.}, \bibinfo{author}{{Marcelli}, L.},
  \bibinfo{author}{{Marsza{\l}}, W.}, \bibinfo{author}{{Mignone}, M.},
  \bibinfo{author}{{Mascetti}, G.}, \bibinfo{author}{{Miyamoto}, H.},
  \bibinfo{author}{{Murashov}, A.}, \bibinfo{author}{{Napolitano}, T.},
  \bibinfo{author}{{Olinto}, A.~V.}, \bibinfo{author}{{Ohmori}, H.},
  \bibinfo{author}{{Osteria}, G.}, \bibinfo{author}{{Panasyuk}, M.},
  \bibinfo{author}{{Porfilio}, M.}, \bibinfo{author}{{Poroshin}, A.},
  \bibinfo{author}{{Parizot}, E.}, \bibinfo{author}{{Picozza}, P.},
  \bibinfo{author}{{Piotrowski}, L.~W.}, \bibinfo{author}{{Plebaniak}, Z.},
  \bibinfo{author}{{Pr{\'e}v{\^o}t}, G.}, \bibinfo{author}{{Przybylak}, M.},
  \bibinfo{author}{{Reali}, E.}, \bibinfo{author}{{Ricci}, M.},
  \bibinfo{author}{{Sakaki}, N.}, \bibinfo{author}{{Shinozaki}, K.},
  \bibinfo{author}{{Szabelski}, J.}, \bibinfo{author}{{Takizawa}, Y.},
  \bibinfo{author}{{Turriziani}, S.}, \bibinfo{author}{{Tra{\"\i}che}, M.},
  \bibinfo{author}{{Valentini}, G.}, \bibinfo{author}{{Wada}, S.},
  \bibinfo{author}{{Wiencke}, L.}, \bibinfo{author}{{Yashin}, I.}, \&
  \bibinfo{author}{{Zuccaro-Marchi}, A.} (\bibinfo{year}{2021}).
\newblock \bibinfo{title}{{Mini-EUSO Mission to Study Earth UV Emissions on
  board the ISS}}.
\newblock {\it \bibinfo{journal}{\apjs}\/},  {\it
  \bibinfo{volume}{253}\/}\bibinfo{issue}{(2)}, \bibinfo{pages}{36}.
  \DOIprefix\doi{10.3847/1538-4365/abd93d}.
  \href{http://arxiv.org/abs/2010.01937}{\tt arXiv:2010.01937}.
\bibitem[{{Benkhoff} et~al.(2010){Benkhoff}, {van Casteren}, {Hayakawa},
  {Fujimoto}, {Laakso}, {Novara}, {Ferri}, {Middleton} \&
  {Ziethe}}]{bepicolombo}
\bibinfo{author}{{Benkhoff}, J.}, \bibinfo{author}{{van Casteren}, J.},
  \bibinfo{author}{{Hayakawa}, H.}, \bibinfo{author}{{Fujimoto}, M.},
  \bibinfo{author}{{Laakso}, H.}, \bibinfo{author}{{Novara}, M.},
  \bibinfo{author}{{Ferri}, P.}, \bibinfo{author}{{Middleton}, H.~R.}, \&
  \bibinfo{author}{{Ziethe}, R.} (\bibinfo{year}{2010}).
\newblock \bibinfo{title}{{BepiColombo{\textemdash}Comprehensive exploration of
  Mercury: Mission overview and science goals}}.
\newblock {\it \bibinfo{journal}{\planss}\/},  {\it
  \bibinfo{volume}{58}\/}\bibinfo{issue}{(1-2)}, \bibinfo{pages}{2--20}.
  \DOIprefix\doi{10.1016/j.pss.2009.09.020}.
\bibitem[{{Benz}(2008)}]{Benz2008}
\bibinfo{author}{{Benz}, A.~O.} (\bibinfo{year}{2008}).
\newblock \bibinfo{title}{{Flare Observations}}.
\newblock {\it \bibinfo{journal}{Living Reviews in Solar Physics}\/},  {\it
  \bibinfo{volume}{5}\/}\bibinfo{issue}{(1)}, \bibinfo{pages}{1}.
  \DOIprefix\doi{10.12942/lrsp-2008-1}.
\bibitem[{{Berrilli} et~al.(2010){Berrilli}, {Bigazzi}, {Roselli}, {Sabatini},
  {Velli}, {Alimenti}, {Cavallini}, {Greco}, {Moretti}, {Orsini}, {Romoli},
  {White}, {ADAHELI Team}, {Ascani}, {Carbone}, {Curti}, {Consolini}, {Di
  Mauro}, {Del Moro}, {Egidi}, {Ermolli}, {Giordano}, {Pastena}, {Pulcino},
  {Pietropaolo}, {Romano}, {Ventura}, {Cauzzi}, {Valdettaro}, {Zuccarello} \&
  {ADAHELI Team}}]{adaheli}
\bibinfo{author}{{Berrilli}, F.}, \bibinfo{author}{{Bigazzi}, A.},
  \bibinfo{author}{{Roselli}, L.}, \bibinfo{author}{{Sabatini}, P.},
  \bibinfo{author}{{Velli}, M.}, \bibinfo{author}{{Alimenti}, F.},
  \bibinfo{author}{{Cavallini}, F.}, \bibinfo{author}{{Greco}, V.},
  \bibinfo{author}{{Moretti}, P.~F.}, \bibinfo{author}{{Orsini}, S.},
  \bibinfo{author}{{Romoli}, M.}, \bibinfo{author}{{White}, S.~M.},
  \bibinfo{author}{{ADAHELI Team}}, \bibinfo{author}{{Ascani}, L.},
  \bibinfo{author}{{Carbone}, V.}, \bibinfo{author}{{Curti}, F.},
  \bibinfo{author}{{Consolini}, G.}, \bibinfo{author}{{Di Mauro}, M.~P.},
  \bibinfo{author}{{Del Moro}, D.}, \bibinfo{author}{{Egidi}, A.},
  \bibinfo{author}{{Ermolli}, I.}, \bibinfo{author}{{Giordano}, S.},
  \bibinfo{author}{{Pastena}, M.}, \bibinfo{author}{{Pulcino}, V.},
  \bibinfo{author}{{Pietropaolo}, E.}, \bibinfo{author}{{Romano}, P.},
  \bibinfo{author}{{Ventura}, P.}, \bibinfo{author}{{Cauzzi}, G.},
  \bibinfo{author}{{Valdettaro}, L.}, \bibinfo{author}{{Zuccarello}, F.}, \&
  \bibinfo{author}{{ADAHELI Team}} (\bibinfo{year}{2010}).
\newblock \bibinfo{title}{{The ADAHELI solar mission: Investigating the
  structure of Sun's lower atmosphere}}.
\newblock {\it \bibinfo{journal}{Advances in Space Research}\/},  {\it
  \bibinfo{volume}{45}\/}\bibinfo{issue}{(10)}, \bibinfo{pages}{1191--1202}.
  \DOIprefix\doi{10.1016/j.asr.2010.01.026}.
\bibitem[{{Berrilli} et~al.(2011){Berrilli}, {Cocciolo}, {Giovannelli}, {Del
  Moro}, {Giannattasio}, {Piazzesi}, {Stangalini}, {Egidi}, {Cavallini},
  {Greco} \& {Selci}}]{fpi}
\bibinfo{author}{{Berrilli}, F.}, \bibinfo{author}{{Cocciolo}, M.},
  \bibinfo{author}{{Giovannelli}, L.}, \bibinfo{author}{{Del Moro}, D.},
  \bibinfo{author}{{Giannattasio}, F.}, \bibinfo{author}{{Piazzesi}, R.},
  \bibinfo{author}{{Stangalini}, M.}, \bibinfo{author}{{Egidi}, A.},
  \bibinfo{author}{{Cavallini}, F.}, \bibinfo{author}{{Greco}, V.}, \&
  \bibinfo{author}{{Selci}, S.} (\bibinfo{year}{2011}).
\newblock \bibinfo{title}{{The Fabry-Perot interferometer prototype for the
  ADAHELI solar small mission}}.
\newblock In \bibinfo{editor}{S.~{Fineschi}}, \&
  \bibinfo{editor}{J.~{Fennelly}} (Eds.), {\it \bibinfo{booktitle}{Solar
  Physics and Space Weather Instrumentation IV}\/} (p.
  \bibinfo{pages}{814807}).
\newblock volume \bibinfo{volume}{8148} of {\it \bibinfo{series}{Society of
  Photo-Optical Instrumentation Engineers (SPIE) Conference Series}\/}.
\newblock \DOIprefix\doi{10.1117/12.893552}.
\bibitem[{{Berrilli} et~al.(2020){Berrilli}, {Criscuoli}, {Penza} \&
  {Lovric}}]{berrilli2020SoPh}
\bibinfo{author}{{Berrilli}, F.}, \bibinfo{author}{{Criscuoli}, S.},
  \bibinfo{author}{{Penza}, V.}, \& \bibinfo{author}{{Lovric}, M.}
  (\bibinfo{year}{2020}).
\newblock \bibinfo{title}{{Long-term (1749-2015) Variations of Solar UV
  Spectral Indices}}.
\newblock {\it \bibinfo{journal}{\solphys}\/},  {\it
  \bibinfo{volume}{295}\/}\bibinfo{issue}{(3)}, \bibinfo{pages}{38}.
  \DOIprefix\doi{10.1007/s11207-020-01603-5}.
\bibitem[{{Berrilli} et~al.(2015){Berrilli}, {Soffitta}, {Velli}, {Sabatini},
  {Bigazzi}, {Bellazzini}, {Bellot Rubio}, {Brez}, {Carbone}, {Cauzzi},
  {Cavallini}, {Consolini}, {Curti}, {Del Moro}, {Di Giorgio}, {Ermolli},
  {Fabiani}, {Faurobert}, {Feller}, {Galsgaard}, {Gburek}, {Giannattasio},
  {Giovannelli}, {Hirzberger}, {Jefferies}, {Madjarska}, {Manni}, {Mazzoni},
  {Muleri}, {Penza}, {Peres}, {Piazzesi}, {Pieralli}, {Pietropaolo}, {Martinez
  Pillet}, {Pinchera}, {Reale}, {Romano}, {Romoli}, {Romoli}, {Rubini},
  {Rudawy}, {Sandri}, {Scardigli}, {Spandre}, {Solanki}, {Stangalini},
  {Vecchio} \& {Zuccarello}}]{adaheli2015}
\bibinfo{author}{{Berrilli}, F.}, \bibinfo{author}{{Soffitta}, P.},
  \bibinfo{author}{{Velli}, M.}, \bibinfo{author}{{Sabatini}, P.},
  \bibinfo{author}{{Bigazzi}, A.}, \bibinfo{author}{{Bellazzini}, R.},
  \bibinfo{author}{{Bellot Rubio}, L.~R.}, \bibinfo{author}{{Brez}, A.},
  \bibinfo{author}{{Carbone}, V.}, \bibinfo{author}{{Cauzzi}, G.},
  \bibinfo{author}{{Cavallini}, F.}, \bibinfo{author}{{Consolini}, G.},
  \bibinfo{author}{{Curti}, F.}, \bibinfo{author}{{Del Moro}, D.},
  \bibinfo{author}{{Di Giorgio}, A.~M.}, \bibinfo{author}{{Ermolli}, I.},
  \bibinfo{author}{{Fabiani}, S.}, \bibinfo{author}{{Faurobert}, M.},
  \bibinfo{author}{{Feller}, A.}, \bibinfo{author}{{Galsgaard}, K.},
  \bibinfo{author}{{Gburek}, S.}, \bibinfo{author}{{Giannattasio}, F.},
  \bibinfo{author}{{Giovannelli}, L.}, \bibinfo{author}{{Hirzberger}, J.},
  \bibinfo{author}{{Jefferies}, S.~M.}, \bibinfo{author}{{Madjarska}, M.~S.},
  \bibinfo{author}{{Manni}, F.}, \bibinfo{author}{{Mazzoni}, A.},
  \bibinfo{author}{{Muleri}, F.}, \bibinfo{author}{{Penza}, V.},
  \bibinfo{author}{{Peres}, G.}, \bibinfo{author}{{Piazzesi}, R.},
  \bibinfo{author}{{Pieralli}, F.}, \bibinfo{author}{{Pietropaolo}, E.},
  \bibinfo{author}{{Martinez Pillet}, V.}, \bibinfo{author}{{Pinchera}, M.},
  \bibinfo{author}{{Reale}, F.}, \bibinfo{author}{{Romano}, P.},
  \bibinfo{author}{{Romoli}, A.}, \bibinfo{author}{{Romoli}, M.},
  \bibinfo{author}{{Rubini}, A.}, \bibinfo{author}{{Rudawy}, P.},
  \bibinfo{author}{{Sandri}, P.}, \bibinfo{author}{{Scardigli}, S.},
  \bibinfo{author}{{Spandre}, G.}, \bibinfo{author}{{Solanki}, S.~K.},
  \bibinfo{author}{{Stangalini}, M.}, \bibinfo{author}{{Vecchio}, A.}, \&
  \bibinfo{author}{{Zuccarello}, F.} (\bibinfo{year}{2015}).
\newblock \bibinfo{title}{{ADAHELI: exploring the fast, dynamic Sun in the
  x-ray, optical, and near-infrared}}.
\newblock {\it \bibinfo{journal}{Journal of Astronomical Telescopes,
  Instruments, and Systems}\/},  {\it \bibinfo{volume}{1}\/},
  \bibinfo{pages}{044006}. \DOIprefix\doi{10.1117/1.JATIS.1.4.044006}.
\bibitem[{{Bigazzi} et~al.(2020){Bigazzi}, {Cauli} \& {Berrilli}}]{bigazzi2020}
\bibinfo{author}{{Bigazzi}, A.}, \bibinfo{author}{{Cauli}, C.}, \&
  \bibinfo{author}{{Berrilli}, F.} (\bibinfo{year}{2020}).
\newblock \bibinfo{title}{{Lower-thermosphere response to solar activity: an
  empirical-mode-decomposition analysis of GOCE 2009-2012 data}}.
\newblock {\it \bibinfo{journal}{Annales Geophysicae}\/},  {\it
  \bibinfo{volume}{38}\/}\bibinfo{issue}{(3)}, \bibinfo{pages}{789--800}.
  \DOIprefix\doi{10.5194/angeo-38-789-2020}.
\bibitem[{{Casolino} et~al.(2021){Casolino}, {Cambie'}, {Marcelli} \&
  {Reali}}]{SiPM}
\bibinfo{author}{{Casolino}, M.}, \bibinfo{author}{{Cambie'}, G.},
  \bibinfo{author}{{Marcelli}, L.}, \& \bibinfo{author}{{Reali}, E.}
  (\bibinfo{year}{2021}).
\newblock \bibinfo{title}{{SiPM development for space-borne and ground
  detectors: From Lazio-Sirad and Mini-EUSO to Lanfos}}.
\newblock {\it \bibinfo{journal}{Nuclear Instruments and Methods in Physics
  Research A}\/},  {\it \bibinfo{volume}{986}\/}, \bibinfo{pages}{164649}.
  \DOIprefix\doi{10.1016/j.nima.2020.164649}.
\bibitem[{{Chupp} et~al.(1973){Chupp}, {Forrest}, {Higbie}, {Suri}, {Tsai} \&
  {Dunphy}}]{chupp1973}
\bibinfo{author}{{Chupp}, E.~L.}, \bibinfo{author}{{Forrest}, D.~J.},
  \bibinfo{author}{{Higbie}, P.~R.}, \bibinfo{author}{{Suri}, A.~N.},
  \bibinfo{author}{{Tsai}, C.}, \& \bibinfo{author}{{Dunphy}, P.~P.}
  (\bibinfo{year}{1973}).
\newblock \bibinfo{title}{{Solar Gamma Ray Lines observed during the Solar
  Activity of August 2 to August 11, 1972}}.
\newblock {\it \bibinfo{journal}{\nat}\/},  {\it
  \bibinfo{volume}{241}\/}\bibinfo{issue}{(5388)}, \bibinfo{pages}{333--335}.
  \DOIprefix\doi{10.1038/241333a0}.
\bibitem[{{De Pontieu} et~al.(2014){De Pontieu}, {Title}, {Lemen}, {Kushner},
  {Akin}, {Allard}, {Berger}, {Boerner}, {Cheung}, {Chou}, {Drake}, {Duncan},
  {Freeland}, {Heyman}, {Hoffman}, {Hurlburt}, {Lindgren}, {Mathur}, {Rehse},
  {Sabolish}, {Seguin}, {Schrijver}, {Tarbell}, {W{\"u}lser}, {Wolfson},
  {Yanari}, {Mudge}, {Nguyen-Phuc}, {Timmons}, {van Bezooijen}, {Weingrod},
  {Brookner}, {Butcher}, {Dougherty}, {Eder}, {Knagenhjelm}, {Larsen},
  {Mansir}, {Phan}, {Boyle}, {Cheimets}, {DeLuca}, {Golub}, {Gates}, {Hertz},
  {McKillop}, {Park}, {Perry}, {Podgorski}, {Reeves}, {Saar}, {Testa}, {Tian},
  {Weber}, {Dunn}, {Eccles}, {Jaeggli}, {Kankelborg}, {Mashburn}, {Pust},
  {Springer}, {Carvalho}, {Kleint}, {Marmie}, {Mazmanian}, {Pereira}, {Sawyer},
  {Strong}, {Worden}, {Carlsson}, {Hansteen}, {Leenaarts}, {Wiesmann},
  {Aloise}, {Chu}, {Bush}, {Scherrer}, {Brekke}, {Martinez-Sykora}, {Lites},
  {McIntosh}, {Uitenbroek}, {Okamoto}, {Gummin}, {Auker}, {Jerram}, {Pool} \&
  {Waltham}}]{IRIS}
\bibinfo{author}{{De Pontieu}, B.}, \bibinfo{author}{{Title}, A.~M.},
  \bibinfo{author}{{Lemen}, J.~R.}, \bibinfo{author}{{Kushner}, G.~D.},
  \bibinfo{author}{{Akin}, D.~J.}, \bibinfo{author}{{Allard}, B.},
  \bibinfo{author}{{Berger}, T.}, \bibinfo{author}{{Boerner}, P.},
  \bibinfo{author}{{Cheung}, M.}, \bibinfo{author}{{Chou}, C.},
  \bibinfo{author}{{Drake}, J.~F.}, \bibinfo{author}{{Duncan}, D.~W.},
  \bibinfo{author}{{Freeland}, S.}, \bibinfo{author}{{Heyman}, G.~F.},
  \bibinfo{author}{{Hoffman}, C.}, \bibinfo{author}{{Hurlburt}, N.~E.},
  \bibinfo{author}{{Lindgren}, R.~W.}, \bibinfo{author}{{Mathur}, D.},
  \bibinfo{author}{{Rehse}, R.}, \bibinfo{author}{{Sabolish}, D.},
  \bibinfo{author}{{Seguin}, R.}, \bibinfo{author}{{Schrijver}, C.~J.},
  \bibinfo{author}{{Tarbell}, T.~D.}, \bibinfo{author}{{W{\"u}lser}, J.~P.},
  \bibinfo{author}{{Wolfson}, C.~J.}, \bibinfo{author}{{Yanari}, C.},
  \bibinfo{author}{{Mudge}, J.}, \bibinfo{author}{{Nguyen-Phuc}, N.},
  \bibinfo{author}{{Timmons}, R.}, \bibinfo{author}{{van Bezooijen}, R.},
  \bibinfo{author}{{Weingrod}, I.}, \bibinfo{author}{{Brookner}, R.},
  \bibinfo{author}{{Butcher}, G.}, \bibinfo{author}{{Dougherty}, B.},
  \bibinfo{author}{{Eder}, J.}, \bibinfo{author}{{Knagenhjelm}, V.},
  \bibinfo{author}{{Larsen}, S.}, \bibinfo{author}{{Mansir}, D.},
  \bibinfo{author}{{Phan}, L.}, \bibinfo{author}{{Boyle}, P.},
  \bibinfo{author}{{Cheimets}, P.~N.}, \bibinfo{author}{{DeLuca}, E.~E.},
  \bibinfo{author}{{Golub}, L.}, \bibinfo{author}{{Gates}, R.},
  \bibinfo{author}{{Hertz}, E.}, \bibinfo{author}{{McKillop}, S.},
  \bibinfo{author}{{Park}, S.}, \bibinfo{author}{{Perry}, T.},
  \bibinfo{author}{{Podgorski}, W.~A.}, \bibinfo{author}{{Reeves}, K.},
  \bibinfo{author}{{Saar}, S.}, \bibinfo{author}{{Testa}, P.},
  \bibinfo{author}{{Tian}, H.}, \bibinfo{author}{{Weber}, M.},
  \bibinfo{author}{{Dunn}, C.}, \bibinfo{author}{{Eccles}, S.},
  \bibinfo{author}{{Jaeggli}, S.~A.}, \bibinfo{author}{{Kankelborg}, C.~C.},
  \bibinfo{author}{{Mashburn}, K.}, \bibinfo{author}{{Pust}, N.},
  \bibinfo{author}{{Springer}, L.}, \bibinfo{author}{{Carvalho}, R.},
  \bibinfo{author}{{Kleint}, L.}, \bibinfo{author}{{Marmie}, J.},
  \bibinfo{author}{{Mazmanian}, E.}, \bibinfo{author}{{Pereira}, T.~M.~D.},
  \bibinfo{author}{{Sawyer}, S.}, \bibinfo{author}{{Strong}, J.},
  \bibinfo{author}{{Worden}, S.~P.}, \bibinfo{author}{{Carlsson}, M.},
  \bibinfo{author}{{Hansteen}, V.~H.}, \bibinfo{author}{{Leenaarts}, J.},
  \bibinfo{author}{{Wiesmann}, M.}, \bibinfo{author}{{Aloise}, J.},
  \bibinfo{author}{{Chu}, K.~C.}, \bibinfo{author}{{Bush}, R.~I.},
  \bibinfo{author}{{Scherrer}, P.~H.}, \bibinfo{author}{{Brekke}, P.},
  \bibinfo{author}{{Martinez-Sykora}, J.}, \bibinfo{author}{{Lites}, B.~W.},
  \bibinfo{author}{{McIntosh}, S.~W.}, \bibinfo{author}{{Uitenbroek}, H.},
  \bibinfo{author}{{Okamoto}, T.~J.}, \bibinfo{author}{{Gummin}, M.~A.},
  \bibinfo{author}{{Auker}, G.}, \bibinfo{author}{{Jerram}, P.},
  \bibinfo{author}{{Pool}, P.}, \& \bibinfo{author}{{Waltham}, N.}
  (\bibinfo{year}{2014}).
\newblock \bibinfo{title}{{The Interface Region Imaging Spectrograph (IRIS)}}.
\newblock {\it \bibinfo{journal}{\solphys}\/},  {\it
  \bibinfo{volume}{289}\/}\bibinfo{issue}{(7)}, \bibinfo{pages}{2733--2779}.
  \DOIprefix\doi{10.1007/s11207-014-0485-y}.
  \href{http://arxiv.org/abs/1401.2491}{\tt arXiv:1401.2491}.
\bibitem[{{Dudok de Wit} et~al.(2009){Dudok de Wit}, {Kretzschmar}, {Lilensten}
  \& {Woods}}]{dewit2009}
\bibinfo{author}{{Dudok de Wit}, T.}, \bibinfo{author}{{Kretzschmar}, M.},
  \bibinfo{author}{{Lilensten}, J.}, \& \bibinfo{author}{{Woods}, T.}
  (\bibinfo{year}{2009}).
\newblock \bibinfo{title}{{Finding the best proxies for the solar UV
  irradiance}}.
\newblock {\it \bibinfo{journal}{\grl}\/},  {\it
  \bibinfo{volume}{36}\/}\bibinfo{issue}{(10)}, \bibinfo{pages}{L10107}.
  \DOIprefix\doi{10.1029/2009GL037825}.
  \href{http://arxiv.org/abs/0905.1811}{\tt arXiv:0905.1811}.
\bibitem[{{Dudok de Wit} \& {Watermann}(2010)}]{dewit2010}
\bibinfo{author}{{Dudok de Wit}, T.}, \& \bibinfo{author}{{Watermann}, J.}
  (\bibinfo{year}{2010}).
\newblock \bibinfo{title}{{Solar forcing of the terrestrial atmosphere}}.
\newblock {\it \bibinfo{journal}{Comptes Rendus Geoscience}\/},  {\it
  \bibinfo{volume}{342}\/}\bibinfo{issue}{(4)}, \bibinfo{pages}{259--272}.
  \DOIprefix\doi{10.1016/j.crte.2009.06.001}.
  \href{http://arxiv.org/abs/0905.1812}{\tt arXiv:0905.1812}.
\bibitem[{{Emslie} et~al.(2005){Emslie}, {Dennis}, {Holman} \&
  {Hudson}}]{emslie2005}
\bibinfo{author}{{Emslie}, A.~G.}, \bibinfo{author}{{Dennis}, B.~R.},
  \bibinfo{author}{{Holman}, G.~D.}, \& \bibinfo{author}{{Hudson}, H.~S.}
  (\bibinfo{year}{2005}).
\newblock \bibinfo{title}{{Refinements to flare energy estimates: A followup to
  ``Energy partition in two solar flare/CME events'' by A. G. Emslie et al.}}
\newblock {\it \bibinfo{journal}{Journal of Geophysical Research (Space
  Physics)}\/},  {\it \bibinfo{volume}{110}\/}\bibinfo{issue}{(A11)},
  \bibinfo{pages}{A11103}. \DOIprefix\doi{10.1029/2005JA011305}.
\bibitem[{{Ermolli} et~al.(2013){Ermolli}, {Matthes}, {Dudok de Wit},
  {Krivova}, {Tourpali}, {Weber}, {Unruh}, {Gray}, {Langematz}, {Pilewskie},
  {Rozanov}, {Schmutz}, {Shapiro}, {Solanki} \& {Woods}}]{ermolli2013}
\bibinfo{author}{{Ermolli}, I.}, \bibinfo{author}{{Matthes}, K.},
  \bibinfo{author}{{Dudok de Wit}, T.}, \bibinfo{author}{{Krivova}, N.~A.},
  \bibinfo{author}{{Tourpali}, K.}, \bibinfo{author}{{Weber}, M.},
  \bibinfo{author}{{Unruh}, Y.~C.}, \bibinfo{author}{{Gray}, L.},
  \bibinfo{author}{{Langematz}, U.}, \bibinfo{author}{{Pilewskie}, P.},
  \bibinfo{author}{{Rozanov}, E.}, \bibinfo{author}{{Schmutz}, W.},
  \bibinfo{author}{{Shapiro}, A.}, \bibinfo{author}{{Solanki}, S.~K.}, \&
  \bibinfo{author}{{Woods}, T.~N.} (\bibinfo{year}{2013}).
\newblock \bibinfo{title}{{Recent variability of the solar spectral irradiance
  and its impact on climate modelling}}.
\newblock {\it \bibinfo{journal}{Atmospheric Chemistry \& Physics}\/},  {\it
  \bibinfo{volume}{13}\/}\bibinfo{issue}{(8)}, \bibinfo{pages}{3945--3977}.
  \DOIprefix\doi{10.5194/acp-13-3945-2013}.
  \href{http://arxiv.org/abs/1303.5577}{\tt arXiv:1303.5577}.
\bibitem[{{Fligge} et~al.(1998){Fligge}, {Solanki}, {Unruh}, {Froehlich} \&
  {Wehrli}}]{SSIFligge1998}
\bibinfo{author}{{Fligge}, M.}, \bibinfo{author}{{Solanki}, S.~K.},
  \bibinfo{author}{{Unruh}, Y.~C.}, \bibinfo{author}{{Froehlich}, C.}, \&
  \bibinfo{author}{{Wehrli}, C.} (\bibinfo{year}{1998}).
\newblock \bibinfo{title}{{A model of solar total and spectral irradiance
  variations}}.
\newblock {\it \bibinfo{journal}{\aap}\/},  {\it \bibinfo{volume}{335}\/},
  \bibinfo{pages}{709--718}.
\bibitem[{{Floyd} et~al.(2002){Floyd}, {Prinz}, {Crane} \&
  {Herring}}]{Floyd2002}
\bibinfo{author}{{Floyd}, L.~E.}, \bibinfo{author}{{Prinz}, D.~K.},
  \bibinfo{author}{{Crane}, P.~C.}, \& \bibinfo{author}{{Herring}, L.~C.}
  (\bibinfo{year}{2002}).
\newblock \bibinfo{title}{{Solar UV irradiance variation during cycles 22 and
  23}}.
\newblock {\it \bibinfo{journal}{Advances in Space Research}\/},  {\it
  \bibinfo{volume}{29}\/}\bibinfo{issue}{(12)}, \bibinfo{pages}{1957--1962}.
  \DOIprefix\doi{10.1016/S0273-1177(02)00242-9}.
\bibitem[{{Fontenla} et~al.(2011){Fontenla}, {Harder}, {Livingston}, {Snow} \&
  {Woods}}]{SSI2011JGRD..11620108F}
\bibinfo{author}{{Fontenla}, J.~M.}, \bibinfo{author}{{Harder}, J.},
  \bibinfo{author}{{Livingston}, W.}, \bibinfo{author}{{Snow}, M.}, \&
  \bibinfo{author}{{Woods}, T.} (\bibinfo{year}{2011}).
\newblock \bibinfo{title}{{High-resolution solar spectral irradiance from
  extreme ultraviolet to far infrared}}.
\newblock {\it \bibinfo{journal}{Journal of Geophysical Research
  (Atmospheres)}\/},  {\it \bibinfo{volume}{116}\/}\bibinfo{issue}{(D20)},
  \bibinfo{pages}{D20108}. \DOIprefix\doi{10.1029/2011JD016032}.
\bibitem[{{Fox} et~al.(2016){Fox}, {Velli}, {Bale}, {Decker}, {Driesman},
  {Howard}, {Kasper}, {Kinnison}, {Kusterer}, {Lario}, {Lockwood}, {McComas},
  {Raouafi} \& {Szabo}}]{PSP}
\bibinfo{author}{{Fox}, N.~J.}, \bibinfo{author}{{Velli}, M.~C.},
  \bibinfo{author}{{Bale}, S.~D.}, \bibinfo{author}{{Decker}, R.},
  \bibinfo{author}{{Driesman}, A.}, \bibinfo{author}{{Howard}, R.~A.},
  \bibinfo{author}{{Kasper}, J.~C.}, \bibinfo{author}{{Kinnison}, J.},
  \bibinfo{author}{{Kusterer}, M.}, \bibinfo{author}{{Lario}, D.},
  \bibinfo{author}{{Lockwood}, M.~K.}, \bibinfo{author}{{McComas}, D.~J.},
  \bibinfo{author}{{Raouafi}, N.~E.}, \& \bibinfo{author}{{Szabo}, A.}
  (\bibinfo{year}{2016}).
\newblock \bibinfo{title}{{The Solar Probe Plus Mission: Humanity's First Visit
  to Our Star}}.
\newblock {\it \bibinfo{journal}{\ssr}\/},  {\it
  \bibinfo{volume}{204}\/}\bibinfo{issue}{(1-4)}, \bibinfo{pages}{7--48}.
  \DOIprefix\doi{10.1007/s11214-015-0211-6}.
\bibitem[{{Giovanelli}(1946)}]{Giovanelli1946}
\bibinfo{author}{{Giovanelli}, R.~G.} (\bibinfo{year}{1946}).
\newblock \bibinfo{title}{{A Theory of Chromospheric Flares}}.
\newblock {\it \bibinfo{journal}{\nat}\/},  {\it
  \bibinfo{volume}{158}\/}\bibinfo{issue}{(4003)}, \bibinfo{pages}{81--82}.
  \DOIprefix\doi{10.1038/158081a0}.
\bibitem[{{Giovannelli} et~al.(2020){Giovannelli}, {Berrilli}, {Calchetti},
  {Del Moro}, {Viavattene}, {Pietropaolo}, {Iarlori}, {Rizi}, {Jefferies},
  {Oliviero}, {Terranegra} \& {Murphy}}]{TSST}
\bibinfo{author}{{Giovannelli}, L.}, \bibinfo{author}{{Berrilli}, F.},
  \bibinfo{author}{{Calchetti}, D.}, \bibinfo{author}{{Del Moro}, D.},
  \bibinfo{author}{{Viavattene}, G.}, \bibinfo{author}{{Pietropaolo}, E.},
  \bibinfo{author}{{Iarlori}, M.}, \bibinfo{author}{{Rizi}, V.},
  \bibinfo{author}{{Jefferies}, S.~M.}, \bibinfo{author}{{Oliviero}, M.},
  \bibinfo{author}{{Terranegra}, L.}, \& \bibinfo{author}{{Murphy}, N.}
  (\bibinfo{year}{2020}).
\newblock \bibinfo{title}{{The Tor Vergata Synoptic Solar Telescope (TSST): A
  robotic, compact facility for solar full disk imaging}}.
\newblock {\it \bibinfo{journal}{Journal of Space Weather and Space
  Climate}\/},  {\it \bibinfo{volume}{10}\/}, \bibinfo{pages}{58}.
  \DOIprefix\doi{10.1051/swsc/2020061}.
\bibitem[{{Gray} et~al.(2010){Gray}, {Beer}, {Geller}, {Haigh}, {Lockwood},
  {Matthes}, {Cubasch}, {Fleitmann}, {Harrison}, {Hood}, {Luterbacher},
  {Meehl}, {Shindell}, {van Geel} \& {White}}]{gray2010}
\bibinfo{author}{{Gray}, L.~J.}, \bibinfo{author}{{Beer}, J.},
  \bibinfo{author}{{Geller}, M.}, \bibinfo{author}{{Haigh}, J.~D.},
  \bibinfo{author}{{Lockwood}, M.}, \bibinfo{author}{{Matthes}, K.},
  \bibinfo{author}{{Cubasch}, U.}, \bibinfo{author}{{Fleitmann}, D.},
  \bibinfo{author}{{Harrison}, G.}, \bibinfo{author}{{Hood}, L.},
  \bibinfo{author}{{Luterbacher}, J.}, \bibinfo{author}{{Meehl}, G.~A.},
  \bibinfo{author}{{Shindell}, D.}, \bibinfo{author}{{van Geel}, B.}, \&
  \bibinfo{author}{{White}, W.} (\bibinfo{year}{2010}).
\newblock \bibinfo{title}{{Solar Influences on Climate}}.
\newblock {\it \bibinfo{journal}{Reviews of Geophysics}\/},  {\it
  \bibinfo{volume}{48}\/}\bibinfo{issue}{(4)}, \bibinfo{pages}{RG4001}.
  \DOIprefix\doi{10.1029/2009RG000282}.
\bibitem[{{Greco} et~al.(2010){Greco}, {Cavallini} \& {Berrilli}}]{isody}
\bibinfo{author}{{Greco}, V.}, \bibinfo{author}{{Cavallini}, F.}, \&
  \bibinfo{author}{{Berrilli}, F.} (\bibinfo{year}{2010}).
\newblock \bibinfo{title}{{The telescope and the double Fabry-P{\'e}rot
  interferometer for the ADAHELI solar space mission}}.
\newblock In \bibinfo{editor}{J.~{Oschmann}, Jacobus~M.},
  \bibinfo{editor}{M.~C. {Clampin}}, \& \bibinfo{editor}{H.~A. {MacEwen}}
  (Eds.), {\it \bibinfo{booktitle}{Space Telescopes and Instrumentation 2010:
  Optical, Infrared, and Millimeter Wave}\/} (p. \bibinfo{pages}{773142}).
\newblock volume \bibinfo{volume}{7731} of {\it \bibinfo{series}{Society of
  Photo-Optical Instrumentation Engineers (SPIE) Conference Series}\/}.
\newblock \DOIprefix\doi{10.1117/12.856617}.
\bibitem[{{Grigis} \& {Benz}(2004)}]{Grigis2004}
\bibinfo{author}{{Grigis}, P.~C.}, \& \bibinfo{author}{{Benz}, A.~O.}
  (\bibinfo{year}{2004}).
\newblock \bibinfo{title}{{The spectral evolution of impulsive solar X-ray
  flares}}.
\newblock {\it \bibinfo{journal}{\aap}\/},  {\it \bibinfo{volume}{426}\/},
  \bibinfo{pages}{1093--1101}. \DOIprefix\doi{10.1051/0004-6361:20041367}.
  \href{http://arxiv.org/abs/astro-ph/0407431}{\tt arXiv:astro-ph/0407431}.
\bibitem[{{Harvey} et~al.(2011){Harvey}, {Bolding}, {Clark}, {Hauth}, {Hill},
  {Kroll}, {Luis}, {Mills}, {Purdy}, {Henney}, {Holland} \& {Winter}}]{gong}
\bibinfo{author}{{Harvey}, J.~W.}, \bibinfo{author}{{Bolding}, J.},
  \bibinfo{author}{{Clark}, R.}, \bibinfo{author}{{Hauth}, D.},
  \bibinfo{author}{{Hill}, F.}, \bibinfo{author}{{Kroll}, R.},
  \bibinfo{author}{{Luis}, G.}, \bibinfo{author}{{Mills}, N.},
  \bibinfo{author}{{Purdy}, T.}, \bibinfo{author}{{Henney}, C.},
  \bibinfo{author}{{Holland}, D.}, \& \bibinfo{author}{{Winter}, J.}
  (\bibinfo{year}{2011}).
\newblock \bibinfo{title}{{Full-disk Solar H-alpha Images From GONG}}.
\newblock In {\it \bibinfo{booktitle}{AAS/Solar Physics Division Abstracts
  \#42}\/} (p. \bibinfo{pages}{17.45}).
\newblock volume~\bibinfo{volume}{42} of {\it \bibinfo{series}{AAS/Solar
  Physics Division Meeting}\/}.
\bibitem[{{Heath} \& {Schlesinger}(1986)}]{heath1986}
\bibinfo{author}{{Heath}, D.~F.}, \& \bibinfo{author}{{Schlesinger}, B.~M.}
  (\bibinfo{year}{1986}).
\newblock \bibinfo{title}{{The Mg 280-nm doublet as a monitor of changes in
  solar ultraviolet irradiance}}.
\newblock {\it \bibinfo{journal}{\jgr}\/},  {\it
  \bibinfo{volume}{91}\/}\bibinfo{issue}{(D8)}, \bibinfo{pages}{8672--8682}.
  \DOIprefix\doi{10.1029/JD091iD08p08672}.
\bibitem[{{Jefferies} et~al.(2006){Jefferies}, {McIntosh}, {Armstrong},
  {Bogdan}, {Cacciani} \& {Fleck}}]{moth}
\bibinfo{author}{{Jefferies}, S.~M.}, \bibinfo{author}{{McIntosh}, S.~W.},
  \bibinfo{author}{{Armstrong}, J.~D.}, \bibinfo{author}{{Bogdan}, T.~J.},
  \bibinfo{author}{{Cacciani}, A.}, \& \bibinfo{author}{{Fleck}, B.}
  (\bibinfo{year}{2006}).
\newblock \bibinfo{title}{{Magnetoacoustic Portals and the Basal Heating of the
  Solar Chromosphere}}.
\newblock {\it \bibinfo{journal}{\apjl}\/},  {\it
  \bibinfo{volume}{648}\/}\bibinfo{issue}{(2)}, \bibinfo{pages}{L151--L155}.
  \DOIprefix\doi{10.1086/508165}.
\bibitem[{{Kerr} et~al.(2015){Kerr}, {Sim{\~o}es}, {Qiu} \&
  {Fletcher}}]{kerr2015}
\bibinfo{author}{{Kerr}, G.~S.}, \bibinfo{author}{{Sim{\~o}es}, P.~J.~A.},
  \bibinfo{author}{{Qiu}, J.}, \& \bibinfo{author}{{Fletcher}, L.}
  (\bibinfo{year}{2015}).
\newblock \bibinfo{title}{{IRIS observations of the Mg ii h and k lines during
  a solar flare}}.
\newblock {\it \bibinfo{journal}{\aap}\/},  {\it \bibinfo{volume}{582}\/},
  \bibinfo{pages}{A50}. \DOIprefix\doi{10.1051/0004-6361/201526128}.
  \href{http://arxiv.org/abs/1508.03813}{\tt arXiv:1508.03813}.
\bibitem[{{Knuth} \& {Glesener}(2020)}]{knuth2020}
\bibinfo{author}{{Knuth}, T.}, \& \bibinfo{author}{{Glesener}, L.}
  (\bibinfo{year}{2020}).
\newblock \bibinfo{title}{{Subsecond Spikes in Fermi GBM X-Ray Flux as a Probe
  for Solar Flare Particle Acceleration}}.
\newblock {\it \bibinfo{journal}{\apj}\/},  {\it
  \bibinfo{volume}{903}\/}\bibinfo{issue}{(1)}, \bibinfo{pages}{63}.
  \DOIprefix\doi{10.3847/1538-4357/abb779}.
  \href{http://arxiv.org/abs/2003.05007}{\tt arXiv:2003.05007}.
\bibitem[{{Kosugi} et~al.(2007){Kosugi}, {Matsuzaki}, {Sakao}, {Shimizu},
  {Sone}, {Tachikawa}, {Hashimoto}, {Minesugi}, {Ohnishi}, {Yamada}, {Tsuneta},
  {Hara}, {Ichimoto}, {Suematsu}, {Shimojo}, {Watanabe}, {Shimada}, {Davis},
  {Hill}, {Owens}, {Title}, {Culhane}, {Harra}, {Doschek} \& {Golub}}]{HINODE}
\bibinfo{author}{{Kosugi}, T.}, \bibinfo{author}{{Matsuzaki}, K.},
  \bibinfo{author}{{Sakao}, T.}, \bibinfo{author}{{Shimizu}, T.},
  \bibinfo{author}{{Sone}, Y.}, \bibinfo{author}{{Tachikawa}, S.},
  \bibinfo{author}{{Hashimoto}, T.}, \bibinfo{author}{{Minesugi}, K.},
  \bibinfo{author}{{Ohnishi}, A.}, \bibinfo{author}{{Yamada}, T.},
  \bibinfo{author}{{Tsuneta}, S.}, \bibinfo{author}{{Hara}, H.},
  \bibinfo{author}{{Ichimoto}, K.}, \bibinfo{author}{{Suematsu}, Y.},
  \bibinfo{author}{{Shimojo}, M.}, \bibinfo{author}{{Watanabe}, T.},
  \bibinfo{author}{{Shimada}, S.}, \bibinfo{author}{{Davis}, J.~M.},
  \bibinfo{author}{{Hill}, L.~D.}, \bibinfo{author}{{Owens}, J.~K.},
  \bibinfo{author}{{Title}, A.~M.}, \bibinfo{author}{{Culhane}, J.~L.},
  \bibinfo{author}{{Harra}, L.~K.}, \bibinfo{author}{{Doschek}, G.~A.}, \&
  \bibinfo{author}{{Golub}, L.} (\bibinfo{year}{2007}).
\newblock \bibinfo{title}{{The Hinode (Solar-B) Mission: An Overview}}.
\newblock {\it \bibinfo{journal}{\solphys}\/},  {\it
  \bibinfo{volume}{243}\/}\bibinfo{issue}{(1)}, \bibinfo{pages}{3--17}.
  \DOIprefix\doi{10.1007/s11207-007-9014-6}.
\bibitem[{{Lean}(1989)}]{lean1989}
\bibinfo{author}{{Lean}, J.} (\bibinfo{year}{1989}).
\newblock \bibinfo{title}{{Contribution of Ultraviolet Irradiance Variations to
  Changes in the Sun's Total Irradiance}}.
\newblock {\it \bibinfo{journal}{Science}\/},  {\it
  \bibinfo{volume}{244}\/}\bibinfo{issue}{(4901)}, \bibinfo{pages}{197--200}.
  \DOIprefix\doi{10.1126/science.244.4901.197}.
\bibitem[{{Lin} et~al.(2002){Lin}, {Dennis}, {Hurford}, {Smith}, {Zehnder},
  {Harvey}, {Curtis}, {Pankow}, {Turin}, {Bester}, {Csillaghy}, {Lewis},
  {Madden}, {van Beek}, {Appleby}, {Raudorf}, {McTiernan}, {Ramaty}, {Schmahl},
  {Schwartz}, {Krucker}, {Abiad}, {Quinn}, {Berg}, {Hashii}, {Sterling},
  {Jackson}, {Pratt}, {Campbell}, {Malone}, {Landis}, {Barrington-Leigh},
  {Slassi-Sennou}, {Cork}, {Clark}, {Amato}, {Orwig}, {Boyle}, {Banks},
  {Shirey}, {Tolbert}, {Zarro}, {Snow}, {Thomsen}, {Henneck}, {McHedlishvili},
  {Ming}, {Fivian}, {Jordan}, {Wanner}, {Crubb}, {Preble}, {Matranga}, {Benz},
  {Hudson}, {Canfield}, {Holman}, {Crannell}, {Kosugi}, {Emslie}, {Vilmer},
  {Brown}, {Johns-Krull}, {Aschwanden}, {Metcalf} \& {Conway}}]{Lin2002}
\bibinfo{author}{{Lin}, R.~P.}, \bibinfo{author}{{Dennis}, B.~R.},
  \bibinfo{author}{{Hurford}, G.~J.}, \bibinfo{author}{{Smith}, D.~M.},
  \bibinfo{author}{{Zehnder}, A.}, \bibinfo{author}{{Harvey}, P.~R.},
  \bibinfo{author}{{Curtis}, D.~W.}, \bibinfo{author}{{Pankow}, D.},
  \bibinfo{author}{{Turin}, P.}, \bibinfo{author}{{Bester}, M.},
  \bibinfo{author}{{Csillaghy}, A.}, \bibinfo{author}{{Lewis}, M.},
  \bibinfo{author}{{Madden}, N.}, \bibinfo{author}{{van Beek}, H.~F.},
  \bibinfo{author}{{Appleby}, M.}, \bibinfo{author}{{Raudorf}, T.},
  \bibinfo{author}{{McTiernan}, J.}, \bibinfo{author}{{Ramaty}, R.},
  \bibinfo{author}{{Schmahl}, E.}, \bibinfo{author}{{Schwartz}, R.},
  \bibinfo{author}{{Krucker}, S.}, \bibinfo{author}{{Abiad}, R.},
  \bibinfo{author}{{Quinn}, T.}, \bibinfo{author}{{Berg}, P.},
  \bibinfo{author}{{Hashii}, M.}, \bibinfo{author}{{Sterling}, R.},
  \bibinfo{author}{{Jackson}, R.}, \bibinfo{author}{{Pratt}, R.},
  \bibinfo{author}{{Campbell}, R.~D.}, \bibinfo{author}{{Malone}, D.},
  \bibinfo{author}{{Landis}, D.}, \bibinfo{author}{{Barrington-Leigh}, C.~P.},
  \bibinfo{author}{{Slassi-Sennou}, S.}, \bibinfo{author}{{Cork}, C.},
  \bibinfo{author}{{Clark}, D.}, \bibinfo{author}{{Amato}, D.},
  \bibinfo{author}{{Orwig}, L.}, \bibinfo{author}{{Boyle}, R.},
  \bibinfo{author}{{Banks}, I.~S.}, \bibinfo{author}{{Shirey}, K.},
  \bibinfo{author}{{Tolbert}, A.~K.}, \bibinfo{author}{{Zarro}, D.},
  \bibinfo{author}{{Snow}, F.}, \bibinfo{author}{{Thomsen}, K.},
  \bibinfo{author}{{Henneck}, R.}, \bibinfo{author}{{McHedlishvili}, A.},
  \bibinfo{author}{{Ming}, P.}, \bibinfo{author}{{Fivian}, M.},
  \bibinfo{author}{{Jordan}, J.}, \bibinfo{author}{{Wanner}, R.},
  \bibinfo{author}{{Crubb}, J.}, \bibinfo{author}{{Preble}, J.},
  \bibinfo{author}{{Matranga}, M.}, \bibinfo{author}{{Benz}, A.},
  \bibinfo{author}{{Hudson}, H.}, \bibinfo{author}{{Canfield}, R.~C.},
  \bibinfo{author}{{Holman}, G.~D.}, \bibinfo{author}{{Crannell}, C.},
  \bibinfo{author}{{Kosugi}, T.}, \bibinfo{author}{{Emslie}, A.~G.},
  \bibinfo{author}{{Vilmer}, N.}, \bibinfo{author}{{Brown}, J.~C.},
  \bibinfo{author}{{Johns-Krull}, C.}, \bibinfo{author}{{Aschwanden}, M.},
  \bibinfo{author}{{Metcalf}, T.}, \& \bibinfo{author}{{Conway}, A.}
  (\bibinfo{year}{2002}).
\newblock \bibinfo{title}{{The Reuven Ramaty High-Energy Solar Spectroscopic
  Imager (RHESSI)}}.
\newblock {\it \bibinfo{journal}{\solphys}\/},  {\it
  \bibinfo{volume}{210}\/}\bibinfo{issue}{(1)}, \bibinfo{pages}{3--32}.
  \DOIprefix\doi{10.1023/A:1022428818870}.
\bibitem[{{Lovric} et~al.(2017){Lovric}, {Tosone}, {Pietropaolo}, {Del Moro},
  {Giovannelli}, {Cagnazzo} \& {Berrilli}}]{lovric2017}
\bibinfo{author}{{Lovric}, M.}, \bibinfo{author}{{Tosone}, F.},
  \bibinfo{author}{{Pietropaolo}, E.}, \bibinfo{author}{{Del Moro}, D.},
  \bibinfo{author}{{Giovannelli}, L.}, \bibinfo{author}{{Cagnazzo}, C.}, \&
  \bibinfo{author}{{Berrilli}, F.} (\bibinfo{year}{2017}).
\newblock \bibinfo{title}{{The dependence of the [FUV-MUV] colour on solar
  cycle}}.
\newblock {\it \bibinfo{journal}{Journal of Space Weather and Space
  Climate}\/},  {\it \bibinfo{volume}{7}\/}, \bibinfo{pages}{A6}.
  \DOIprefix\doi{10.1051/swsc/2017001}.
  \href{http://arxiv.org/abs/1606.08267}{\tt arXiv:1606.08267}.
\bibitem[{{Matsuoka} et~al.(2009){Matsuoka}, {Kawasaki}, {Ueno}, {Tomida},
  {Kohama}, {Suzuki}, {Adachi}, {Ishikawa}, {Mihara}, {Sugizaki}, {Isobe},
  {Nakagawa}, {Tsunemi}, {Miyata}, {Kawai}, {Kataoka}, {Morii}, {Yoshida},
  {Negoro}, {Nakajima}, {Ueda}, {Chujo}, {Yamaoka}, {Yamazaki}, {Nakahira},
  {You}, {Ishiwata}, {Miyoshi}, {Eguchi}, {Hiroi}, {Katayama} \&
  {Ebisawa}}]{MAXI}
\bibinfo{author}{{Matsuoka}, M.}, \bibinfo{author}{{Kawasaki}, K.},
  \bibinfo{author}{{Ueno}, S.}, \bibinfo{author}{{Tomida}, H.},
  \bibinfo{author}{{Kohama}, M.}, \bibinfo{author}{{Suzuki}, M.},
  \bibinfo{author}{{Adachi}, Y.}, \bibinfo{author}{{Ishikawa}, M.},
  \bibinfo{author}{{Mihara}, T.}, \bibinfo{author}{{Sugizaki}, M.},
  \bibinfo{author}{{Isobe}, N.}, \bibinfo{author}{{Nakagawa}, Y.},
  \bibinfo{author}{{Tsunemi}, H.}, \bibinfo{author}{{Miyata}, E.},
  \bibinfo{author}{{Kawai}, N.}, \bibinfo{author}{{Kataoka}, J.},
  \bibinfo{author}{{Morii}, M.}, \bibinfo{author}{{Yoshida}, A.},
  \bibinfo{author}{{Negoro}, H.}, \bibinfo{author}{{Nakajima}, M.},
  \bibinfo{author}{{Ueda}, Y.}, \bibinfo{author}{{Chujo}, H.},
  \bibinfo{author}{{Yamaoka}, K.}, \bibinfo{author}{{Yamazaki}, O.},
  \bibinfo{author}{{Nakahira}, S.}, \bibinfo{author}{{You}, T.},
  \bibinfo{author}{{Ishiwata}, R.}, \bibinfo{author}{{Miyoshi}, S.},
  \bibinfo{author}{{Eguchi}, S.}, \bibinfo{author}{{Hiroi}, K.},
  \bibinfo{author}{{Katayama}, H.}, \& \bibinfo{author}{{Ebisawa}, K.}
  (\bibinfo{year}{2009}).
\newblock \bibinfo{title}{{The MAXI Mission on the ISS: Science and Instruments
  for Monitoring All-Sky X-Ray Images}}.
\newblock {\it \bibinfo{journal}{\pasj}\/},  {\it \bibinfo{volume}{61}\/},
  \bibinfo{pages}{999}. \DOIprefix\doi{10.1093/pasj/61.5.999}.
  \href{http://arxiv.org/abs/0906.0631}{\tt arXiv:0906.0631}.
\bibitem[{{Meier}(1991)}]{Meier1991}
\bibinfo{author}{{Meier}, R.~R.} (\bibinfo{year}{1991}).
\newblock \bibinfo{title}{{Ultraviolet spectroscopy and remote sensing of the
  upper atmosphere}}.
\newblock {\it \bibinfo{journal}{\ssr}\/},  {\it
  \bibinfo{volume}{58}\/}\bibinfo{issue}{(1)}, \bibinfo{pages}{1--185}.
  \DOIprefix\doi{10.1007/BF01206000}.
\bibitem[{{Milbourne} et~al.(2019){Milbourne}, {Haywood}, {Phillips}, {Saar},
  {Cegla}, {Cameron}, {Costes}, {Dumusque}, {Langellier}, {Latham},
  {Maldonado}, {Malavolta}, {Mortier}, {Palumbo}, {Thompson}, {Watson},
  {Bouchy}, {Buchschacher}, {Cecconi}, {Charbonneau}, {Cosentino}, {Ghedina},
  {Glenday}, {Gonzalez}, {Li}, {Lodi}, {L{\'o}pez-Morales}, {Lovis}, {Mayor},
  {Micela}, {Molinari}, {Pepe}, {Piotto}, {Rice}, {Sasselov}, {S{\'e}gransan},
  {Sozzetti}, {Szentgyorgyi}, {Udry} \& {Walsworth}}]{Milbourne2019}
\bibinfo{author}{{Milbourne}, T.~W.}, \bibinfo{author}{{Haywood}, R.~D.},
  \bibinfo{author}{{Phillips}, D.~F.}, \bibinfo{author}{{Saar}, S.~H.},
  \bibinfo{author}{{Cegla}, H.~M.}, \bibinfo{author}{{Cameron}, A.~C.},
  \bibinfo{author}{{Costes}, J.}, \bibinfo{author}{{Dumusque}, X.},
  \bibinfo{author}{{Langellier}, N.}, \bibinfo{author}{{Latham}, D.~W.},
  \bibinfo{author}{{Maldonado}, J.}, \bibinfo{author}{{Malavolta}, L.},
  \bibinfo{author}{{Mortier}, A.}, \bibinfo{author}{{Palumbo}, I., M.~L.},
  \bibinfo{author}{{Thompson}, S.}, \bibinfo{author}{{Watson}, C.~A.},
  \bibinfo{author}{{Bouchy}, F.}, \bibinfo{author}{{Buchschacher}, N.},
  \bibinfo{author}{{Cecconi}, M.}, \bibinfo{author}{{Charbonneau}, D.},
  \bibinfo{author}{{Cosentino}, R.}, \bibinfo{author}{{Ghedina}, A.},
  \bibinfo{author}{{Glenday}, A.~G.}, \bibinfo{author}{{Gonzalez}, M.},
  \bibinfo{author}{{Li}, C.~H.}, \bibinfo{author}{{Lodi}, M.},
  \bibinfo{author}{{L{\'o}pez-Morales}, M.}, \bibinfo{author}{{Lovis}, C.},
  \bibinfo{author}{{Mayor}, M.}, \bibinfo{author}{{Micela}, G.},
  \bibinfo{author}{{Molinari}, E.}, \bibinfo{author}{{Pepe}, F.},
  \bibinfo{author}{{Piotto}, G.}, \bibinfo{author}{{Rice}, K.},
  \bibinfo{author}{{Sasselov}, D.}, \bibinfo{author}{{S{\'e}gransan}, D.},
  \bibinfo{author}{{Sozzetti}, A.}, \bibinfo{author}{{Szentgyorgyi}, A.},
  \bibinfo{author}{{Udry}, S.}, \& \bibinfo{author}{{Walsworth}, R.~L.}
  (\bibinfo{year}{2019}).
\newblock \bibinfo{title}{{HARPS-N Solar RVs Are Dominated by Large, Bright
  Magnetic Regions}}.
\newblock {\it \bibinfo{journal}{\apj}\/},  {\it
  \bibinfo{volume}{874}\/}\bibinfo{issue}{(1)}, \bibinfo{pages}{107}.
  \DOIprefix\doi{10.3847/1538-4357/ab064a}.
  \href{http://arxiv.org/abs/1902.04184}{\tt arXiv:1902.04184}.
\bibitem[{{M{\"u}ller} et~al.(2020){M{\"u}ller}, {St. Cyr}, {Zouganelis},
  {Gilbert}, {Marsden}, {Nieves-Chinchilla}, {Antonucci}, {Auch{\`e}re},
  {Berghmans}, {Horbury}, {Howard}, {Krucker}, {Maksimovic}, {Owen}, {Rochus},
  {Rodriguez-Pacheco}, {Romoli}, {Solanki}, {Bruno}, {Carlsson}, {Fludra},
  {Harra}, {Hassler}, {Livi}, {Louarn}, {Peter}, {Sch{\"u}hle}, {Teriaca}, {del
  Toro Iniesta}, {Wimmer-Schweingruber}, {Marsch}, {Velli}, {De Groof}, {Walsh}
  \& {Williams}}]{SO}
\bibinfo{author}{{M{\"u}ller}, D.}, \bibinfo{author}{{St. Cyr}, O.~C.},
  \bibinfo{author}{{Zouganelis}, I.}, \bibinfo{author}{{Gilbert}, H.~R.},
  \bibinfo{author}{{Marsden}, R.}, \bibinfo{author}{{Nieves-Chinchilla}, T.},
  \bibinfo{author}{{Antonucci}, E.}, \bibinfo{author}{{Auch{\`e}re}, F.},
  \bibinfo{author}{{Berghmans}, D.}, \bibinfo{author}{{Horbury}, T.~S.},
  \bibinfo{author}{{Howard}, R.~A.}, \bibinfo{author}{{Krucker}, S.},
  \bibinfo{author}{{Maksimovic}, M.}, \bibinfo{author}{{Owen}, C.~J.},
  \bibinfo{author}{{Rochus}, P.}, \bibinfo{author}{{Rodriguez-Pacheco}, J.},
  \bibinfo{author}{{Romoli}, M.}, \bibinfo{author}{{Solanki}, S.~K.},
  \bibinfo{author}{{Bruno}, R.}, \bibinfo{author}{{Carlsson}, M.},
  \bibinfo{author}{{Fludra}, A.}, \bibinfo{author}{{Harra}, L.},
  \bibinfo{author}{{Hassler}, D.~M.}, \bibinfo{author}{{Livi}, S.},
  \bibinfo{author}{{Louarn}, P.}, \bibinfo{author}{{Peter}, H.},
  \bibinfo{author}{{Sch{\"u}hle}, U.}, \bibinfo{author}{{Teriaca}, L.},
  \bibinfo{author}{{del Toro Iniesta}, J.~C.},
  \bibinfo{author}{{Wimmer-Schweingruber}, R.~F.}, \bibinfo{author}{{Marsch},
  E.}, \bibinfo{author}{{Velli}, M.}, \bibinfo{author}{{De Groof}, A.},
  \bibinfo{author}{{Walsh}, A.}, \& \bibinfo{author}{{Williams}, D.}
  (\bibinfo{year}{2020}).
\newblock \bibinfo{title}{{The Solar Orbiter mission. Science overview}}.
\newblock {\it \bibinfo{journal}{\aap}\/},  {\it \bibinfo{volume}{642}\/},
  \bibinfo{pages}{A1}. \DOIprefix\doi{10.1051/0004-6361/202038467}.
  \href{http://arxiv.org/abs/2009.00861}{\tt arXiv:2009.00861}.
\bibitem[{{Muralikrishna} et~al.(2022){Muralikrishna}, {dos Santos} \&
  {Vieira}}]{SSI2022JSWSC}
\bibinfo{author}{{Muralikrishna}, A.}, \bibinfo{author}{{dos Santos}, R.
  D.~C.}, \& \bibinfo{author}{{Vieira}, L. E.~A.} (\bibinfo{year}{2022}).
\newblock \bibinfo{title}{{Exploring possibilities for solar irradiance
  prediction from solar photosphere images using recurrent neural networks}}.
\newblock {\it \bibinfo{journal}{Journal of Space Weather and Space
  Climate}\/},  {\it \bibinfo{volume}{12}\/}, \bibinfo{pages}{19}.
  \DOIprefix\doi{10.1051/swsc/2022015}.
\bibitem[{Narici et~al.(2018)Narici, Rizzo, Berrilli \& Del~Moro}]{narici2018}
\bibinfo{author}{Narici, L.}, \bibinfo{author}{Rizzo, A.},
  \bibinfo{author}{Berrilli, F.}, \& \bibinfo{author}{Del~Moro, D.}
  (\bibinfo{year}{2018}).
\newblock \bibinfo{title}{Solar particle events and human deep space
  exploration: Measurements and considerations}.
\newblock In {\it \bibinfo{booktitle}{Extreme Events in Geospace}\/} (pp.
  \bibinfo{pages}{433--451}).
\newblock \bibinfo{publisher}{Elsevier}.
\bibitem[{{Orozco Suárez} et~al.(2022){Orozco Suárez}, {del Toro Iniesta},
  {Bailén}, {López Jiménez}, {Balaguez Jiménez}, {Ishikawa}, {Katsukawa},
  {Kano}, {Shimizu}, {Trujillo Bueno}, {Asensio Ramos} \& {del Pino
  Alemán}}]{Orozco2022}
\bibinfo{author}{{Orozco Suárez}, D.}, \bibinfo{author}{{del Toro Iniesta},
  J.~C.}, \bibinfo{author}{{Bailén}, F.~J.}, \bibinfo{author}{{López
  Jiménez}, A.}, \bibinfo{author}{{Balaguez Jiménez}, L.~R., M.~and{Bellot
  Rubio}}, \bibinfo{author}{{Ishikawa}, R.}, \bibinfo{author}{{Katsukawa}, Y.},
  \bibinfo{author}{{Kano}, R.}, \bibinfo{author}{{Shimizu}, T.},
  \bibinfo{author}{{Trujillo Bueno}, J.}, \bibinfo{author}{{Asensio Ramos},
  A.}, \& \bibinfo{author}{{del Pino Alemán}, T.} (\bibinfo{year}{2022}).
\newblock \bibinfo{title}{{CASPER: A mission to study the time-dependent
  evolution of the magnetic solar chromosphere and transition regions}}.
\newblock {\it \bibinfo{journal}{Exp. Astr.}\/}, .
  \DOIprefix\doi{10.1007/s10686-022-09839-8}.
\bibitem[{{Penza} et~al.(2021){Penza}, {Berrilli}, {Bertello}, {Cantoresi} \&
  {Criscuoli}}]{Penza2021}
\bibinfo{author}{{Penza}, V.}, \bibinfo{author}{{Berrilli}, F.},
  \bibinfo{author}{{Bertello}, L.}, \bibinfo{author}{{Cantoresi}, M.}, \&
  \bibinfo{author}{{Criscuoli}, S.} (\bibinfo{year}{2021}).
\newblock \bibinfo{title}{{Prediction of Sunspot and Plage Coverage for Solar
  Cycle 25}}.
\newblock {\it \bibinfo{journal}{\apjl}\/},  {\it
  \bibinfo{volume}{922}\/}\bibinfo{issue}{(1)}, \bibinfo{pages}{L12}.
  \DOIprefix\doi{10.3847/2041-8213/ac3663}.
  \href{http://arxiv.org/abs/2111.02928}{\tt arXiv:2111.02928}.
\bibitem[{{Pesnell} et~al.(2012){Pesnell}, {Thompson} \& {Chamberlin}}]{SDO}
\bibinfo{author}{{Pesnell}, W.~D.}, \bibinfo{author}{{Thompson}, B.~J.}, \&
  \bibinfo{author}{{Chamberlin}, P.~C.} (\bibinfo{year}{2012}).
\newblock \bibinfo{title}{{The Solar Dynamics Observatory (SDO)}}.
\newblock {\it \bibinfo{journal}{\solphys}\/},  {\it
  \bibinfo{volume}{275}\/}\bibinfo{issue}{(1-2)}, \bibinfo{pages}{3--15}.
  \DOIprefix\doi{10.1007/s11207-011-9841-3}.
\bibitem[{Piana et~al.(2022)Piana, Emslie, Massone \& Dennis}]{piana2022}
\bibinfo{author}{Piana, M.}, \bibinfo{author}{Emslie, A.~G.},
  \bibinfo{author}{Massone, A.~M.}, \& \bibinfo{author}{Dennis, B.~R.}
  (\bibinfo{year}{2022}).
\newblock {\it \bibinfo{title}{Hard X-Ray Imaging of Solar Flares}\/}.
\newblock \bibinfo{publisher}{Springer}.
\bibitem[{{Plainaki} et~al.(2020){Plainaki}, {Antonucci}, {Bemporad},
  {Berrilli}, {Bertucci}, {Castronuovo}, {De Michelis}, {Giardino}, {Iuppa},
  {Laurenza}, {Marcucci}, {Messerotti}, {Narici}, {Negri}, {Nozzoli}, {Orsini},
  {Romano}, {Cavallini}, {Polenta} \& {Ippolito}}]{plainaki2020}
\bibinfo{author}{{Plainaki}, C.}, \bibinfo{author}{{Antonucci}, M.},
  \bibinfo{author}{{Bemporad}, A.}, \bibinfo{author}{{Berrilli}, F.},
  \bibinfo{author}{{Bertucci}, B.}, \bibinfo{author}{{Castronuovo}, M.},
  \bibinfo{author}{{De Michelis}, P.}, \bibinfo{author}{{Giardino}, M.},
  \bibinfo{author}{{Iuppa}, R.}, \bibinfo{author}{{Laurenza}, M.},
  \bibinfo{author}{{Marcucci}, F.}, \bibinfo{author}{{Messerotti}, M.},
  \bibinfo{author}{{Narici}, L.}, \bibinfo{author}{{Negri}, B.},
  \bibinfo{author}{{Nozzoli}, F.}, \bibinfo{author}{{Orsini}, S.},
  \bibinfo{author}{{Romano}, V.}, \bibinfo{author}{{Cavallini}, E.},
  \bibinfo{author}{{Polenta}, G.}, \& \bibinfo{author}{{Ippolito}, A.}
  (\bibinfo{year}{2020}).
\newblock \bibinfo{title}{{Current state and perspectives of Space Weather
  science in Italy}}.
\newblock {\it \bibinfo{journal}{Journal of Space Weather and Space
  Climate}\/},  {\it \bibinfo{volume}{10}\/}, \bibinfo{pages}{6}.
  \DOIprefix\doi{10.1051/swsc/2020003}.
\bibitem[{{Poghosyan} \& {Golkar}(2017)}]{cubesat}
\bibinfo{author}{{Poghosyan}, A.}, \& \bibinfo{author}{{Golkar}, A.}
  (\bibinfo{year}{2017}).
\newblock \bibinfo{title}{{CubeSat evolution: Analyzing CubeSat capabilities
  for conducting science missions}}.
\newblock {\it \bibinfo{journal}{Progress in Aerospace Sciences}\/},  {\it
  \bibinfo{volume}{88}\/}, \bibinfo{pages}{59--83}.
  \DOIprefix\doi{10.1016/j.paerosci.2016.11.002}.
\bibitem[{{P{\"o}tzi} et~al.(2021){P{\"o}tzi}, {Veronig}, {Jarolim},
  {Rodr{\'\i}guez G{\'o}mez}, {Podlachikova}, {Baumgartner}, {Freislich} \&
  {Strutzmann}}]{graz}
\bibinfo{author}{{P{\"o}tzi}, W.}, \bibinfo{author}{{Veronig}, A.},
  \bibinfo{author}{{Jarolim}, R.}, \bibinfo{author}{{Rodr{\'\i}guez G{\'o}mez},
  J.~M.}, \bibinfo{author}{{Podlachikova}, T.}, \bibinfo{author}{{Baumgartner},
  D.}, \bibinfo{author}{{Freislich}, H.}, \& \bibinfo{author}{{Strutzmann}, H.}
  (\bibinfo{year}{2021}).
\newblock \bibinfo{title}{{Kanzelh{\"o}he Observatory: Instruments, Data
  Processing and Data Products}}.
\newblock {\it \bibinfo{journal}{\solphys}\/},  {\it
  \bibinfo{volume}{296}\/}\bibinfo{issue}{(11)}, \bibinfo{pages}{164}.
  \DOIprefix\doi{10.1007/s11207-021-01903-4}.
  \href{http://arxiv.org/abs/2111.03176}{\tt arXiv:2111.03176}.
\bibitem[{{Ryan} et~al.(2012){Ryan}, {Milligan}, {Gallagher}, {Dennis},
  {Tolbert}, {Schwartz} \& {Young}}]{GOES}
\bibinfo{author}{{Ryan}, D.~F.}, \bibinfo{author}{{Milligan}, R.~O.},
  \bibinfo{author}{{Gallagher}, P.~T.}, \bibinfo{author}{{Dennis}, B.~R.},
  \bibinfo{author}{{Tolbert}, A.~K.}, \bibinfo{author}{{Schwartz}, R.~A.}, \&
  \bibinfo{author}{{Young}, C.~A.} (\bibinfo{year}{2012}).
\newblock \bibinfo{title}{{The Thermal Properties of Solar Flares over Three
  Solar Cycles Using GOES X-Ray Observations}}.
\newblock {\it \bibinfo{journal}{\apjs}\/},  {\it
  \bibinfo{volume}{202}\/}\bibinfo{issue}{(2)}, \bibinfo{pages}{11}.
  \DOIprefix\doi{10.1088/0067-0049/202/2/11}.
  \href{http://arxiv.org/abs/1206.1005}{\tt arXiv:1206.1005}.
\bibitem[{{Shibata} \& {Tanuma}(2001)}]{Shibata2001}
\bibinfo{author}{{Shibata}, K.}, \& \bibinfo{author}{{Tanuma}, S.}
  (\bibinfo{year}{2001}).
\newblock \bibinfo{title}{{Plasmoid-induced-reconnection and fractal
  reconnection}}.
\newblock {\it \bibinfo{journal}{Earth, Planets and Space}\/},  {\it
  \bibinfo{volume}{53}\/}, \bibinfo{pages}{473--482}.
  \DOIprefix\doi{10.1186/BF03353258}.
  \href{http://arxiv.org/abs/astro-ph/0101008}{\tt arXiv:astro-ph/0101008}.
\bibitem[{{Smith} \& {Scalo}(2007)}]{smith2007}
\bibinfo{author}{{Smith}, D.~S.}, \& \bibinfo{author}{{Scalo}, J.~M.}
  (\bibinfo{year}{2007}).
\newblock \bibinfo{title}{{Risks due to X-ray flares during astronaut
  extravehicular activity}}.
\newblock {\it \bibinfo{journal}{Space Weather}\/},  {\it
  \bibinfo{volume}{5}\/}\bibinfo{issue}{(6)}, \bibinfo{pages}{S06004}.
  \DOIprefix\doi{10.1029/2006SW000300}.
  \href{http://arxiv.org/abs/astro-ph/0701314}{\tt arXiv:astro-ph/0701314}.
\bibitem[{{Snow} et~al.(2010){Snow}, {McClintock} \& {Woods}}]{snow2010}
\bibinfo{author}{{Snow}, M.}, \bibinfo{author}{{McClintock}, W.~E.}, \&
  \bibinfo{author}{{Woods}, T.~N.} (\bibinfo{year}{2010}).
\newblock \bibinfo{title}{{Solar spectral irradiance variability in the
  ultraviolet from SORCE and UARS SOLSTICE}}.
\newblock {\it \bibinfo{journal}{Advances in Space Research}\/},  {\it
  \bibinfo{volume}{46}\/}\bibinfo{issue}{(3)}, \bibinfo{pages}{296--302}.
  \DOIprefix\doi{10.1016/j.asr.2010.03.027}.
\bibitem[{{Snow} et~al.(2005){Snow}, {McClintock}, {Woods}, {White}, {Harder}
  \& {Rottman}}]{snow2005}
\bibinfo{author}{{Snow}, M.}, \bibinfo{author}{{McClintock}, W.~E.},
  \bibinfo{author}{{Woods}, T.~N.}, \bibinfo{author}{{White}, O.~R.},
  \bibinfo{author}{{Harder}, J.~W.}, \& \bibinfo{author}{{Rottman}, G.}
  (\bibinfo{year}{2005}).
\newblock \bibinfo{title}{{The Mg II Index from SORCE}}.
\newblock {\it \bibinfo{journal}{\solphys}\/},  {\it
  \bibinfo{volume}{230}\/}\bibinfo{issue}{(1-2)}, \bibinfo{pages}{325--344}.
  \DOIprefix\doi{10.1007/s11207-005-6879-0}.
\bibitem[{{Szenicer} et~al.(2019){Szenicer}, {Fouhey}, {Munoz-Jaramillo},
  {Wright}, {Thomas}, {Galvez}, {Jin} \& {Cheung}}]{SSI2019SciA....5.6548S}
\bibinfo{author}{{Szenicer}, A.}, \bibinfo{author}{{Fouhey}, D.~F.},
  \bibinfo{author}{{Munoz-Jaramillo}, A.}, \bibinfo{author}{{Wright}, P.~J.},
  \bibinfo{author}{{Thomas}, R.}, \bibinfo{author}{{Galvez}, R.},
  \bibinfo{author}{{Jin}, M.}, \& \bibinfo{author}{{Cheung}, M. C.~M.}
  (\bibinfo{year}{2019}).
\newblock \bibinfo{title}{{A deep learning virtual instrument for monitoring
  extreme UV solar spectral irradiance}}.
\newblock {\it \bibinfo{journal}{Science Advances}\/},  {\it
  \bibinfo{volume}{5}\/}\bibinfo{issue}{(10)}, \bibinfo{pages}{eaaw6548}.
  \DOIprefix\doi{10.1126/sciadv.aaw6548}.
\bibitem[{{Tripathi} et~al.(2017){Tripathi}, {Ramaprakash}, {Khan}, {Ghosh},
  {Chatterjee}, {Banerjee}, {Chordia}, {Gandorfer}, {Krivova}, {Nandy},
  {Rajarshi} \& {Solanki}}]{aditya}
\bibinfo{author}{{Tripathi}, D.}, \bibinfo{author}{{Ramaprakash}, A.~N.},
  \bibinfo{author}{{Khan}, A.}, \bibinfo{author}{{Ghosh}, A.},
  \bibinfo{author}{{Chatterjee}, S.}, \bibinfo{author}{{Banerjee}, D.},
  \bibinfo{author}{{Chordia}, P.}, \bibinfo{author}{{Gandorfer}, A.},
  \bibinfo{author}{{Krivova}, N.}, \bibinfo{author}{{Nandy}, D.},
  \bibinfo{author}{{Rajarshi}, C.}, \& \bibinfo{author}{{Solanki}, S.~K.}
  (\bibinfo{year}{2017}).
\newblock \bibinfo{title}{{The Solar Ultraviolet Imaging Telescope on-board
  Aditya-L1}}.
\newblock {\it \bibinfo{journal}{Current Science}\/},  {\it
  \bibinfo{volume}{113}\/}\bibinfo{issue}{(4)}, \bibinfo{pages}{616}.
  \DOIprefix\doi{10.18520/cs/v113/i04/616-619}.
\bibitem[{{Unruh} et~al.(2008){Unruh}, {Krivova}, {Solanki}, {Harder} \&
  {Kopp}}]{SSI2008A&A...486..311U}
\bibinfo{author}{{Unruh}, Y.~C.}, \bibinfo{author}{{Krivova}, N.~A.},
  \bibinfo{author}{{Solanki}, S.~K.}, \bibinfo{author}{{Harder}, J.~W.}, \&
  \bibinfo{author}{{Kopp}, G.} (\bibinfo{year}{2008}).
\newblock \bibinfo{title}{{Spectral irradiance variations: comparison between
  observations and the SATIRE model on solar rotation time scales}}.
\newblock {\it \bibinfo{journal}{\aap}\/},  {\it
  \bibinfo{volume}{486}\/}\bibinfo{issue}{(1)}, \bibinfo{pages}{311--323}.
  \DOIprefix\doi{10.1051/0004-6361:20078421}.
  \href{http://arxiv.org/abs/0802.4178}{\tt arXiv:0802.4178}.
\bibitem[{{Viereck} \& {Puga}(1999)}]{viereck1999}
\bibinfo{author}{{Viereck}, R.~A.}, \& \bibinfo{author}{{Puga}, L.~C.}
  (\bibinfo{year}{1999}).
\newblock \bibinfo{title}{{The NOAA Mg II core-to-wing solar index:
  Construction of a 20-year time series of chromospheric variability from
  multiple satellites}}.
\newblock {\it \bibinfo{journal}{\jgr}\/},  {\it
  \bibinfo{volume}{104}\/}\bibinfo{issue}{(A5)}, \bibinfo{pages}{9995--10006}.
  \DOIprefix\doi{10.1029/1998JA900163}.
\bibitem[{{White} et~al.(1998){White}, {Livingston}, {Keil} \&
  {Henry}}]{white1998}
\bibinfo{author}{{White}, O.~R.}, \bibinfo{author}{{Livingston}, W.~C.},
  \bibinfo{author}{{Keil}, S.~L.}, \& \bibinfo{author}{{Henry}, T.~W.}
  (\bibinfo{year}{1998}).
\newblock \bibinfo{title}{{Variability of the Solar Call K Line over the 22
  Year Hale Cycle}}.
\newblock In \bibinfo{editor}{K.~S. {Balasubramaniam}},
  \bibinfo{editor}{J.~{Harvey}}, \& \bibinfo{editor}{D.~{Rabin}} (Eds.), {\it
  \bibinfo{booktitle}{Synoptic Solar Physics}\/} (p. \bibinfo{pages}{293}).
\newblock volume \bibinfo{volume}{140} of {\it \bibinfo{series}{Astronomical
  Society of the Pacific Conference Series}\/}.
\bibitem[{{Woods} et~al.(2004){Woods}, {Eparvier}, {Fontenla}, {Harder},
  {Kopp}, {McClintock}, {Rottman}, {Smiley} \& {Snow}}]{woods2004}
\bibinfo{author}{{Woods}, T.~N.}, \bibinfo{author}{{Eparvier}, F.~G.},
  \bibinfo{author}{{Fontenla}, J.}, \bibinfo{author}{{Harder}, J.},
  \bibinfo{author}{{Kopp}, G.}, \bibinfo{author}{{McClintock}, W.~E.},
  \bibinfo{author}{{Rottman}, G.}, \bibinfo{author}{{Smiley}, B.}, \&
  \bibinfo{author}{{Snow}, M.} (\bibinfo{year}{2004}).
\newblock \bibinfo{title}{{Solar irradiance variability during the October 2003
  solar storm period}}.
\newblock {\it \bibinfo{journal}{\grl}\/},  {\it
  \bibinfo{volume}{31}\/}\bibinfo{issue}{(10)}, \bibinfo{pages}{L10802}.
  \DOIprefix\doi{10.1029/2004GL019571}.
\bibitem[{{Woods} et~al.(2021){Woods}, {Harder}, {Kopp}, {McCabe}, {Rottman},
  {Ryan} \& {Snow}}]{woods2021}
\bibinfo{author}{{Woods}, T.~N.}, \bibinfo{author}{{Harder}, J.~W.},
  \bibinfo{author}{{Kopp}, G.}, \bibinfo{author}{{McCabe}, D.},
  \bibinfo{author}{{Rottman}, G.}, \bibinfo{author}{{Ryan}, S.}, \&
  \bibinfo{author}{{Snow}, M.} (\bibinfo{year}{2021}).
\newblock \bibinfo{title}{{Overview of the Solar Radiation and Climate
  Experiment (SORCE) Seventeen-Year Mission}}.
\newblock {\it \bibinfo{journal}{\solphys}\/},  {\it
  \bibinfo{volume}{296}\/}\bibinfo{issue}{(8)}, \bibinfo{pages}{127}.
  \DOIprefix\doi{10.1007/s11207-021-01869-3}.

\end{thebibliography}

\end{document}